\documentstyle[prd,aps,eqsecnum,preprint,tighten,floats,psfig]{revtex}

\def\ha{{1\over 2}}

\def\psibar{\overline{\psi}}
\def\del{\partial}

\def\g{\gamma}

\begin{document}
\draft

\preprint{\vbox{\hfill SLAC-PUB-8890 \\
          \vbox{\hfill UMN-D-01-4}  \\
          \vbox{\hfill SMUHEP/01-07 }
          \vbox{\vskip0.3in}
          }}

\title{Application of Pauli--Villars regularization 
and discretized light-cone quantization 
to a single-fermion truncation of Yukawa theory%
\footnote{\baselineskip=14pt
Work supported in part by the U.S. Department of Energy,
contracts DE-AC03-76SF00515, DE-FG02-98ER41087,
and DE-FG03-95ER40908.}}

\author{Stanley J. Brodsky}
\address{Stanford Linear Accelerator Center,
Stanford University, Stanford, California 94309}

\author{John R. Hiller}
\address{Department of Physics,
University of Minnesota-Duluth, Duluth, Minnesota 55812}

\author{Gary McCartor}
\address{Department of Physics,
Southern Methodist University, Dallas, Texas 75275}

\date{\today}

\maketitle

\begin{abstract}
We apply Pauli--Villars regularization and
discretized light-cone quantization to the nonperturbative
solution of (3+1)-dimensional Yukawa theory in a single-fermion
truncation.  Three heavy scalars, including two
with negative norm, are used to regulate the theory.
The matrix eigenvalue problem is solved for the
lowest-mass state with use of a new, indefinite-metric
Lanczos algorithm.  Various observables are extracted
from the wave functions, including
average multiplicities and average momenta of
constituents, structure functions, and a
form factor slope.
\end{abstract}
\pacs{12.38.Lg,11.15.Tk,11.10.Gh,02.60.Nm
\begin{center}(To appear in Physical Review D.)\end{center}}

\narrowtext

%%%%%%%%%%%%%%%%%%%%%%
\section{Introduction}
%%%%%%%%%%%%%%%%%%%%%%

Light-cone Hamiltonian diagonalization methods
offer a number of attractive advantages for solving nonperturbative
problems in quantum field theory, such as a physical Minkowski space
description, boost invariance of the bound-state wave functions, 
no requirement for fermion doubling, and a consistent 
Fock-state expansion well matched to physical problems.  
In the discretized light-cone quantization (DLCQ) method, 
the light-cone Hamiltonian $H_{\rm LC}$
of a quantum field theory is diagonalized on a discrete Fock basis defined
by assuming periodic boundary conditions in the light-cone
coordinates~\cite{PauliBrodsky,DLCQreview}. The eigenvalues of $H_{\rm LC}$
give the mass spectrum of the theory, and the respective eigenfunctions
projected on the free Fock basis provide the frame-independent light-cone
wave functions needed for phenomenology~\cite{Brodsky:2001jw} including the
amplitudes needed to compute exclusive $B$ decays~\cite{Beneke:2000ry,Keum:2001wi,Brodsky:1999hn},
deeply virtual Compton scattering~\cite{Brodsky:2001xy,Diehl:2001xz},
and other hard exclusive processes~\cite{Brodsky:2001dx}. 
The DLCQ method has been successfully used to solve a large number of
one-space and one-time theories~\cite{DLCQreview}, including supersymmetric
gauge theories~\cite{Lunin:2001im}. It also has found application in
analyzing confinement mechanisms~\cite{Gross:1998mx}, string
theory~\cite{Pinskyetal}, and $M$-theory~\cite{Antonuccio:1997cf}.

The application of DLCQ to physical, (3+1)-dimensional space-time
quantum field theories is computationally challenging because of the rapid
growth of the number of degrees of freedom as the size of the Fock
representation grows.  A promising alternative is the transverse lattice
method~\cite{Dalley} which combines light-cone methods in the longitudinal 
light-cone direction with a space-time lattice for the transverse dimensions.
Recently Dalley~\cite{Dalley:2001gj} and Burkhart and Seal~\cite{Seal:2000gs} 
have extended the transverse lattice method to estimate
the shape of the valence light-cone wave function of a pion, a key input to
much hadron phenomenology.  
Burkhart and Seal have also given an explicit calculation of the 
Isgur--Wise function for semileptonic $B$ decays~\cite{Burkardt:2001dy}.

Another major difficulty in applying DLCQ to quantum field theory in 3+1
dimensions is the implementation of a nonperturbative renormalization
method.  Most methods of regulating nonperturbative calculations in the
light-cone representation, such as momentum cutoffs, do not allow a correct
renormalization even of perturbative calculations. The problem can be
traced to the fact that any momentum cutoff violates Lorentz invariance as
well as gauge invariance~\cite{PV1}. Since dimensional regularization is
not available in DLCQ, one needs to introduce new fields or degrees of
freedom to render the ultraviolet behavior of the theory finite.  One
intriguing possibility is to analyze ultraviolet-finite supersymmetric
theories and then introduce breaking of the theory.  The heavy
supersymmetric partners then regulate the ordinary sector of the theory in
a manner analogous to Pauli--Villars (PV) regulation~\cite{PauliVillars}. String
theory also provides mechanisms for regulating quantum field theory at
short distances which are equivalent to an infinite spectrum of
PV particles~\cite{Dienes:2001se}. The introduction of
PV fields can thus regulate a theory covariantly, after which
the discretized momentum grid of DLCQ acts only as a numerical tool in the
manner of performing a numerical integral.

In our previous work~\cite{PV1,PV2} we have shown that a model field
theory in 3+1 dimensions can be solved by combining
DLCQ with PV regulation of the ultraviolet regime.  In our first
application~\cite{PV1}, a model theory was constructed to have an
exact analytic solution by which the DLCQ results could be checked, for
both accuracy and rapidity of convergence.  The model was regulated in the
ultraviolet by a single PV boson, which was included in the
DLCQ Fock basis in the same way as the ``physical'' particles of the
theory.  We then extended this approach to a more realistic model which
mimics many features of a full quantum field theory~\cite{PV2}. Unlike the
analytic model which contained a static source, the light-cone energies of
the particles in this model have the correct longitudinal and transverse
momentum dependence.

An important question is whether the generalized PV method with a 
finite number of fields can regulate a field theory at all orders.
Paston and Franke~\cite{Paston:1997hs} have studied the relation between
perturbation theory in the light-cone representation and standard Feynman
perturbation theory, and they have developed general rules for testing
regularization procedures.  For full Yukawa theory, Paston, Franke and
Prokhvatilov~\cite{Paston:1999er} have shown that one PV boson
and two PV fermions can regulate the theory in such a way as to
allow a correct perturbative renormalization.

In this paper we shall apply generalized PV regularization and
discrete light-cone quantization to the nonperturbative solution of 
(3+1)-dimensional Yukawa theory in a single-fermion truncation.  
We allow any number of
bosons in the Yukawa theory but only one fermion in the Fock
representation; fermion pair terms and any other terms that involve
anti-fermions are neglected.  We shall thus consider a field-theoretic
model where one particle, which we take to be a fermion of mass $M$, acts
as a dynamical source and sink for bosons of mass $\mu$. In addition,
three heavy PV scalars, including two with negative norm, will
be used to regulate the theory so that the chiral properties of the
renormalized theory are maintained, at least to one loop in perturbation
theory.  In particular, the mass of the renormalized fermion constituent
vanishes as its bare mass vanishes.  A distinct advantage of our approach
is that the counterterms are generated automatically by the PV
particles and their negative-metric couplings.
We emphasize that the PV fields are added from the beginning,
at the Lagrangian level, to facilitate our nonperturbative
calculations, rather than being invoked for subtractions
at a diagrammatic level, which would limit the implementation
to perturbation theory.
Note that the PV fields enter singly, doubly or triply at all three-point
vertices. This contrasts with use of generalized Pauli--Villars spectral
integrals over mass to regulate divergences.

The extra degrees of freedom of the PV scalars and their
negative-metric couplings introduce new computational
challenges.  However, the matrix eigenvalue problem can be solved for the
lowest-mass state by the use of a new, indefinite-metric Lanczos algorithm
which we describe in an Appendix.  We also calculate the light-cone
wave function of each Fock-sector component and the values for various
physical quantities, such as average multiplicities and average momenta of
constituents, bosonic and fermionic structure functions, and a form factor
slope.  We also verify that with our choice of
PV conditions, the DLCQ calculations of the
nonperturbative theory at weak coupling coincide with the covariant
perturbation theory through one loop, although numerical resolution does
start to become a problem even on a supercomputer.

In our convention, we define light-cone coordinates~\cite{Dirac} by
\begin{equation}
x^\pm = x^0+x^3\,,\;\;
\bbox{x}_\perp=(x^1,x^2)\,.
\end{equation}
The time coordinate is taken to be $x^+$.  The dot
product of two four-vectors is
\begin{equation}
p\cdot x=\frac{1}{2}(p^+x^- + p^-x^+)
                -\bbox{p}_\perp\cdot\bbox{x}_\perp\,.
\end{equation}
Thus the momentum component conjugate to $x^-$ is $p^+$,
and the light-cone energy is $p^-$.  We use underscores
to identify light-cone three-vectors, such as
\begin{equation}
\underline{p}=(p^+,\bbox{p}_\perp)\,.
\end{equation}
For additional details, see Appendix A of Ref.~\cite{PV1}
or a review paper~\cite{DLCQreview}.

The following is an outline of the remainder of the paper.
In Sec.~\ref{sec:Yukawa} we discuss the regularization
and renormalization of the Yukawa Hamiltonian.  Our
numerical methods and results are presented in Sec.~\ref{sec:Results}.
Section~\ref{sec:Conclusions} contains our conclusions
and plans for future work.  Details of the numerical
diagonalization method are given in the Appendix.

%%%%%%%%%%%%%%%%%%%%%%%%%%
\section{Yukawa theory}  \label{sec:Yukawa}
%%%%%%%%%%%%%%%%%%%%%%%%%%

\subsection{Light-cone Hamiltonian}

We write the Lagrangian for Yukawa theory as
\begin{equation}
{\cal L}=\ha(\del_\mu\phi)^2-\ha\mu^2\phi^2
	+{i\over2}\Bigl[\psibar\g^\mu\del_\mu
          -(\del_\mu\psibar)\g^\mu\Bigr]\psi
		-M\psibar\psi - g\phi\psibar\psi\,.
\end{equation}
The corresponding light-cone Hamiltonian has been given
by McCartor and Robertson~\cite{McCartorRobertson}.  Here we
consider a single-fermion truncation and therefore neglect
pair terms and any other terms that involve anti-fermions.
We also neglect longitudinal zero modes.  To regulate the
theory, we include three PV bosons.  The resulting
Hamiltonian is 
\begin{eqnarray} \label{eq:HLC}
\lefteqn{H_{\rm LC}=
   \sum_{\underline{n},s}
      \frac{M^2+\delta M^2+({\bbox n}_\perp \pi/L_\perp)^2}{n/K}
          b_{\underline{n},s}^\dagger b_{\underline{n},s}
   +\sum_{\underline{m}i}
          \frac{\mu_i^2+({\bbox m}_\perp \pi/L_\perp)^2}{m/K}(-1)^i
              a_{i\underline{m}}^\dagger a_{i\underline{m}}} \nonumber \\
   && +\frac{g\sqrt{\pi}}{2L_\perp^2}
          \sum_{\underline{n}\underline{m}}\sum_{si}\frac{\xi_i}{\sqrt{m}}
     \left(\left[\frac{{\bbox \epsilon}_{-2s}^{\,*}\cdot{\bbox n}_\perp}{n/K}
         +\frac{{\bbox \epsilon}_{2s}\cdot({\bbox n}_\perp+{\bbox m}_\perp)}
                                               {(n+m)/K}        \right]
     b_{\underline{n}+\underline{m},-s}^\dagger b_{\underline{n},s} 
                                            a_{i\underline{m}} 
          + \mbox{h.c.}\right) \nonumber\\
   && +\frac{Mg}{\sqrt{8\pi}L_\perp}
             \sum_{\underline{n}\underline{m}}\sum_{si}\frac{\xi_i}{\sqrt{m}}
     \left(\left[\frac{1}{n/K}+\frac{1}{(n+m)/K}\right]
     b_{\underline{n}+\underline{m},s}^\dagger b_{\underline{n},s} 
                a_{i\underline{m}}    + \mbox{h.c.} \right)  \\
   && +\frac{g^2}{8\pi L_\perp^2}
              \sum_{\underline{n}\underline{m}\underline{m}'}\sum_{sij}
         \frac{\xi_i\xi_j}{\sqrt{mm'}}
     \left[\left(b_{\underline{n}+\underline{m}+\underline{m}',s}^\dagger 
         b_{\underline{n},s} a_{i\underline{m}'} a_{j\underline{m}}
                  \frac{1}{(n+m)/K}+ \mbox{h.c.} \right)\right. \nonumber \\
    &&\rule{0.25in}{0mm} \left.
        + b_{\underline{n}+\underline{m}-\underline{m}',s}^\dagger 
            b_{\underline{n},s} a_{i\underline{m}'}^\dagger a_{j\underline{m}}
                  \left(\frac{1}{(n-m')/K}+\frac{1}{(n+m)/K}\right)\right]\,,
  \nonumber
\end{eqnarray}
where $M$ is the fermion mass, 
$\mu\equiv\mu_0$ is the physical boson mass, 
$\mu_i$ and $(-1)^i$ are the mass and norm of the i-th PV boson, and
${\bbox \epsilon}_{2s}\equiv-\frac{1}{\sqrt{2}}(2s,i)$.
The nonzero commutators are 
\begin{equation}
\left[a_{i\underline{m}},a_{j\underline{m}'}^\dagger\right]
          =(-1)^i\delta_{ij}\delta_{\underline{m},\underline{m}'}\,, \;\;
   \left\{b_{\underline{n},s},b_{\underline{n}',s'}^\dagger\right\}
     =\delta_{\underline{n},\underline{n}'} \delta_{s,s'}\,.   
\end{equation}
The different boson couplings are denoted by $\xi_i g$, where
$\xi_0\equiv 1$ corresponds to the physical boson; the other $\xi_i$
are chosen to arrange cancellations discussed below.
Fermion self-induced inertia terms cancel in a sum over
bosons and therefore do not appear.
The bare parameters are the coupling $g$ and the mass shift $\delta M^2$.

The number of PV flavors is determined by the
cancellations needed to regulate the theory and restore 
chiral invariance in the $M=0$ limit~\cite{ChangYan,PV1}.
The discrete chiral symmetry $\psi\rightarrow i\gamma_5\psi$, 
$\phi\rightarrow -\phi$ should itself guarantee a zero self-mass 
in this limit, if not for the fact that the symmetry is typically 
broken by a cutoff.  
The one-loop self-energy of the fermion is~\cite{PV1}
\begin{eqnarray}  \label{eq:Iapprox}
I(\mu^2,M^2,\Lambda^2)&\approx& \frac{\alpha}{2\pi}
     \Biggl[ \left( \frac{\Lambda^2}{2} - \mu^2 \ln \Lambda^2 
           + \mu^2 \ln \mu^2 - \frac{\mu^4}{2\Lambda^2}\right)
  \nonumber \\
&& + M^2\left( 3 \ln\Lambda^2 - 3 \ln \mu^2 - \frac{9}{2}
                           + \frac{5\mu^2}{\Lambda^2}\right)
   \\
&& + M^4\left(\frac{2}{\mu^2}\ln (M^2/\mu^2)
                 +\frac{1}{3\mu^2}-\frac{1}{2\Lambda^2}\right)
                           \Biggr]\,,
\nonumber
\end{eqnarray}
with $\Lambda^2$ a cutoff such that $\Lambda^2\gg \mu^2\gg M^2$.
In order that the self-energy be finite and proportional to $M^2$,
the relative coupling strengths $\xi_i$ must
satisfy the constraints
\begin{equation}\label{eq:constraints}
1+\sum_{i=1}^3 (-1)^i\xi_i^2=0\,, \;\;
\mu^2+\sum_{i=1}^3 (-1)^i\xi_i^2\mu_i^2=0\,, \;\;
\sum_{i=1}^3 (-1)^i\xi_i^2\mu_i^2\ln(\mu_i^2/\mu^2)=0\,.
\end{equation}
In addition, the norm of the i-th PV field must be chosen as $(-1)^i$.

The third constraint in (\ref{eq:constraints}) is peculiar 
to the one-loop calculation.
For higher-order or nonperturbative calculations, it must be
replaced by a more general renormalization condition.  However,
to simplify the numerical work, we use the one-loop constraint
and then check for failure of cancellation in the zero-mass limit.

As it stands, $H_{\rm LC}$ contains infrared divergences
associated with the instantaneous fermion interaction,
which is singular at the point where the instantaneous
fermion has zero longitudinal momentum.
The divergences are partly cancelled by crossed-boson contributions.  
To cancel the remainder we need
to add an effective interaction,
modeled on the missing fermion Z graph.  
The effective interaction
is constructed from the pair creation
and annihilation terms in the Yukawa light-cone energy 
operator~\cite{McCartorRobertson}
\begin{eqnarray}
{\cal P}_{\rm pair}^-&&= \frac{g}{2L_\perp\sqrt{L}}
   \sum_{\underline{p}\underline{q}si}
  \left[\frac{{\bbox \epsilon}_{-2s}\cdot{\bbox p}_\perp}{p^+\sqrt{q^+}}
        +\frac{{\bbox \epsilon}_{2s}^{\,*}\cdot({\bbox q}_\perp-{\bbox p}_\perp)}
                    {(q^+-p^+)\sqrt{q^+}}\right]\xi_i
     b_{\underline{p},s}^\dagger d_{\underline{q}-\underline{p},s}^\dagger 
                               a_{i\underline{q}}    +\mbox{h.c.} \\
   && +\frac{Mg}{2L_\perp\sqrt{2L}}
   \sum_{\underline{p}\underline{q}si}
    \left[\frac{1}{p^+\sqrt{q^+}}-\frac{1}{(q^+-p^+)\sqrt{q^+}}\right]\xi_i
     b_{\underline{p},s}^\dagger d_{\underline{q}-\underline{p},-s}^\dagger 
               a_{i\underline{q}}     +\mbox{h.c.}  \nonumber
\end{eqnarray}
The denominator for the intermediate state is
\begin{equation}
\frac{M^2}{P^+}-p_{\rm spectators}^- -\frac{M^2+p_\perp^{\prime 2}}{p'^+}
         -\frac{M^2+({\bbox q}_\perp^{\,\prime}-{\bbox p}_\perp)^2}{q'^+-p^+}
          -\frac{M^2+p_\perp^2}{p^+}\,. 
\end{equation}
To guarantee the cancellation of the singularity in the numerical
calculation, the instantaneous
interaction is kept only if the corresponding crossed-boson graph 
is permitted by the numerical cutoffs.

The Fock-state expansion for the single-fermion eigenstate 
of the Hamiltonian is
\begin{eqnarray}
\Phi_\sigma&=&\sqrt{16\pi^3P^+}\sum_{n_0,n_1,n_2,n_3=0}^\infty
     \int\frac{dp^+d^2p_\perp}{\sqrt{16\pi^3p^+}}
   \prod_{j=1}^{n_{\rm tot}}
     \int\frac{dq_j^+d^2q_{\perp j}}{\sqrt{16\pi^3q_j^+}} 
        \sum_s\\
   &  & \times \delta(\underline{P}-\underline{p}
                     -\sum_{j}^{n_{\rm tot}}\underline{q}_j)
     \phi_{\sigma s}^{(n_i)}(\underline{q}_j;\underline{p})
     \frac{1}{\sqrt{\prod_i n_i!}}b_{\underline{p}s}^\dagger
      \prod_j^{n_{\rm tot}} a_{i_j\underline{q}_j}^\dagger |0\rangle \,,
\nonumber
\end{eqnarray}
where $n_0$ is the number of physical bosons, $n_i$ the
number of PV bosons of flavor $i$, $n_{\rm tot}=\sum_{i=0}^3 n_i$,
and $i_j$ is the flavor of the j-th constituent boson.  It solves
the eigenvalue problem $H_{\rm LC}\Phi_\sigma=M^2\Phi_\sigma$.
The normalization is
\begin{equation}
\Phi_\sigma^{\prime\dagger}\cdot\Phi_\sigma
=16\pi^3P^+\delta(\underline{P}'-\underline{P})\,. 
\end{equation}

\subsection{Renormalization conditions}

Mass renormalization is carried out by rearranging the eigenvalue problem 
into one for $\delta M^2$ at fixed $M^2$
\begin{eqnarray}
x\left[M^2-
   \frac{M^2+p_\perp^2}{x}\right.
   &-&\left.\sum_i\frac{\mu_i^2+q_{\perp i}^2}{y_i}\right] 
                                   \tilde{\phi}(\underline{q}_i)     \nonumber \\
&-&\int\prod_j dy'_j d^2q'_{\perp j}\sqrt{xx'}
{\cal K}(\underline{q}_i,\underline{q}'_j)\tilde{\phi}'(\underline{q}'_j)=
                             \delta M^2\tilde{\phi}(\underline{q}_i)\,.
\end{eqnarray}
Here $x=p^+/P^+$ is the fermion momentum fraction,
${\cal K}$ is shorthand for the interaction kernel,
and the $\tilde{\phi}\equiv\phi/\sqrt{x}$ are new wave functions.

The coupling is fixed by setting a value for the expectation value
$\langle :\!\!\phi^2(0)\!\!:\rangle
\equiv\Phi_\sigma^\dagger\!:\!\!\phi^2(0)\!\!:\!\Phi_\sigma$
for the boson field operator $\phi$.
This quantity is useful because it can be computed fairly efficiently in a sum
similar to a normalization sum
\begin{eqnarray}
\langle :\!\!\phi^2(0)\!\!:\rangle
        &=&\sum_{n_i=0}^\infty \int \prod_j^{n_{\rm tot}}
                          \,dq_j^+d^2q_{\perp j} \sum_s (-1)^{(n_i)}
   \\
  & & \times \left(\sum_{k=1}^n \frac{2}{q_k^+/P^+}\right)
              \left|\phi_{\sigma s}^{(n_i)}(\underline{q}_j;
       \underline{P}-\sum_j\underline{q}_j)\right|^2\,,
\nonumber
\end{eqnarray}
with $(-1)^{(n_i)}$ being the norm of the state with 
boson flavor partitioning $(n_i)$.

%%%%%%%%%%%%%%%%%%%%%%%%%%%%%%%%%%%%%%%%%%%%%%%%%%%%%%%%%%%%%%%%%%%
\section{Results} \label{sec:Results}
%%%%%%%%%%%%%%%%%%%%%%%%%%%%%%%%%%%%%%%%%%%%%%%%%%%%%%%%%%%%%%%%%%%

\subsection{Numerical methods}

The principal tools for the solution of the Hamiltonian
eigenvalue problem are DLCQ~\cite{DLCQreview} and the
Lanczos diagonalization algorithm~\cite{Lanczos}.  The
first converts the problem to a matrix form, and the second
quickly generates one or more eigenvectors.  The use of
negatively-normed states makes the ordinary Lanczos algorithm
inapplicable; however, an efficient generalization has
been developed for this situation.  It is discussed in
the Appendix.  The constraint for the coupling renormalization
is solved iteratively.

The discretization is based on the standard DLCQ approach where
longitudinal and transverse momenta are assigned discrete values
$q^+=m\pi/L$ and ${\bbox q}_\perp={\bbox n}_\perp \pi/L_\perp$, 
with $L$ and $L_\perp$ chosen length scales.  Momentum conservation 
requires the individual $m$ to sum to a fixed constant $K$, where $P^+=K\pi/L$
is the total longitudinal momentum.  The integer $K$ is called
the harmonic resolution~\cite{PauliBrodsky}, because longitudinal
momentum fractions $y=q^+/P^+=m/K$ are resolved to order 1/K.
The positivity of longitudinal momenta forces a natural cutoff such
that $m\leq K$.  Also, the eigenvalue problem for $M^2$ is independent
of $L$; the length scale cancels between $P^+$ and $P^-$.

The boundary conditions in the longitudinal direction are chosen
to be periodic for the boson fields but antiperiodic for the fermion field.
This means that the integers $m$ are even for bosons, and the corresponding fermion
momentum index is odd.  The harmonic resolution $K$ is then also odd
for the single-fermion state considered here.

The transverse direction requires an explicit cutoff $\Lambda^2$, which
we impose on individual light-cone energies
\begin{equation}
(\mu^2+m_\perp^2\pi^2/L_\perp^2)/m \leq \Lambda^2/K\,.
\end{equation}
The total transverse momentum is taken to be zero.  The integers $n_x$
and $n_y$ are limited to a range $[-N_\perp,N_\perp]$, which, along with the 
cutoff, determines the transverse scale $L_\perp$.

Given this discretization, the eigenvalue problem is converted
to a matrix problem by a trapezoidal approximation to any momentum
integrals.  We have found useful modifications~\cite{PV1,PV2}
which include non-constant weighting factors near the integral
boundaries.  These weights are adjusted to compensate for the
DLCQ grid being incommensurate with the boundaries.  In the
present calculation, these weights are kept only in the two-body
sector where maximal symmetry can be maintained.  For higher
Fock sectors, sensitivity to cancellations is of greater
concern than boundary effects, and the weights disrupt the
cancellations.

Unlike an ordinary eigenvalue problem, the presence of negatively
normed states allows unphysical states to be lower in mass
than the physical one-fermion state.  Criteria must be employed
to select the correct state in a numerical calculation.
We used the following: a positive norm, a real eigenvalue, 
a minimum number of nodes (preferably zero) in the parallel-helicity
boson-fermion wave function,
and the largest bare-fermion probability between 0 and 1.
Each of these characteristics can be computed without constructing 
the full eigenvector, provided one saves a few components of each
Lanczos vector ${\bbox q}_j$ (see the Appendix). 

As a check on the calculation, we took advantage of an exact 
solution that exists for the unphysical situation of equal-mass
PV bosons~\cite{PV4}.  In a particular (null) basis the matrix
representation is purely triangular.  Each wave function of
the dressed fermion is then directly computable in a finite
number of steps.

For comparison, we also solved the problem using Brillouin--Wigner
perturbation theory, for which
\begin{equation}
\Phi_\sigma \simeq \sqrt{Z}\sum_{n=0} 
    \left(\frac{1-b_{\underline{K}}^\dagger|0 \rangle\langle0|b_{\underline{K}}}
          {\delta M^2-{\cal K}_{\rm diag}}
            {\cal K}_{\rm off-diag}\right)^n 
               b_{\underline{K}}^\dagger|0 \rangle\,. 
\end{equation}
The integrals were computed numerically with the same discretization
as the nonperturbative DLCQ calculation.  The main effort in the
perturbative expansion is then matrix multiplication, just as
for the Lanczos algorithm.

\subsection{Computed quantities}

Various quantities can be computed from the
wave functions $\phi^{(n_i)}$ for the
different Fock sectors.  We compute the slope of the no-flip form
factor of the fermion, structure functions for bosons and the fermion,
average momenta, average multiplicities, and a quantity sensitive to boson
correlations. The form factor slope $F'(0)$ is given by~\cite{PV1}
\begin{eqnarray} \label{eq:BetterFprime}
F'(0)&=&-\sum_{n_i=0}^\infty\int\,\prod_j dq_j^+d^2q_{\perp j}
           \sum_s (-1)^{(n_i)}   \\
     && \times \left[\sum_{k=1}^{n_{\rm tot}} \left|\frac{y_k}{2}\nabla_{\perp k}
             \phi_{\sigma s}^{(n_i)}(\underline{q}_j;
                                       \underline{P}-\sum_j\underline{q}_j)
                                     \right|^2 \right]\,.
    \nonumber
\end{eqnarray}
The physical boson structure function for bare helicity $s$ is defined as
\begin{eqnarray}
f_{Bs}(y)&\equiv&\sum_{n_i=0}^\infty\int\,\prod_jdq_j^+d^2q_{\perp j}
       \sum_{k=1}^{n_0}\delta(y-q_k^+/P^+) (-1)^{(n_i)}\\
   &&\times \left|\phi_{\sigma s}^{(n_i)}(\underline{q}_j;
                      \underline{P}-\sum_i\underline{q}_j)\right|^2\,.
   \nonumber
\end{eqnarray}
The normalization is such that the integral yields the average 
multiplicity $\langle n_B\rangle$.  We separate two pieces,
for parallel and antiparallel bare helicity
\begin{equation}
\langle n_{Bs}\rangle=\int_0^1f_{Bs}(y)dy\,,
\end{equation}
and have
\begin{equation}
\langle n_B\rangle=\langle n_{B+}\rangle+\langle n_{B-}\rangle\,.
\end{equation}
The average boson momentum is treated analogously, with
\begin{equation}
\langle y_{Bs}\rangle=\int_0^1yf_{Bs}(y)dy\;\;
\mbox{and} \;\;
\langle y_B\rangle=\langle y_{B+}\rangle+\langle y_{B-}\rangle\,.
\end{equation}
As a measure of the correlations in the multiple-boson Fock sectors,
we compute
$(\langle y_1 y_2\rangle-\langle y\rangle^2)_{n\geq 2}/\langle y\rangle^2$
where
\begin{eqnarray}
\langle y_1 y_2\rangle_{n\geq 2}&=
  &\sum_{n_0\geq2,n_i}\int\,\prod_jdq_j^+d^2q_{\perp j} \sum_s
       \sum_{k_1\neq k_2}^{n_0} \frac{q_{k_1}^+}{P^+}
                                     \frac{q_{k_2}^+}{P^+} (-1)^{(n_i)} \\
   &&\times \left|\phi_{\sigma s}^{(n_i)}(\underline{q}_j;
                      \underline{P}-\sum_j\underline{q}_j)\right|^2\,,
   \nonumber
\end{eqnarray}
and $\langle y\rangle_{n\geq 2}$ is the same as $\langle y\rangle$
except that only states with two or more bosons are included.

\begin{figure}[htbp]
\centerline{\psfig{file=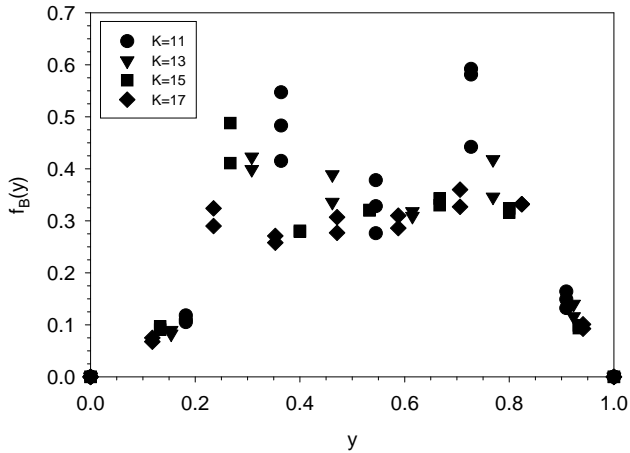,width=3.5in}}
\caption{\label{fig:badfB}
The boson structure function $f_B$ at various low to moderate
numerical resolutions, with $M=\mu$, %\protect\\
$\langle:\!\!\phi^2(0)\!\!:\rangle=1.0$,
$\Lambda^2=50\mu^2$, $\mu_1^2=10\mu^2$, $\mu_2^2=20\mu^2$, and
$\mu_3^2=30\mu^2$.  To compare with a structure function computed
at higher resolutions, see Fig.~\ref{fig:fBL50}(d).} 
\end{figure}

Calculations at low resolutions tend to have difficulty for stronger
coupling.  This can be seen already at third order in perturbation
theory, where loop integrals are poorly approximated and subtractions
between loops with different boson flavors magnify the errors.  Fully
averaged quantities such as $\langle n_B\rangle$ are less affected
by this, but the structure function $f_B$ can be quite poorly represented.
An example is given in Fig.~\ref{fig:badfB}.

Clearly one cure is to work at higher resolution.  Limitations
on computer storage then require truncation in the number of
constituents.  Luckily, even at the strongest coupling that
we consider, states with large numbers of constituents are
unimportant, and yet the results differ significantly from
lowest-order perturbation theory.  The relative importance
of different numbers of constituents can be seen in 
Table~\ref{tab:FSprobmax4}, where we list probabilities
for the various Fock sector contributions to a calculation
with moderate resolution.  Truncation to a maximum of two
bosons is seen to offer a very good approximation.
For weaker couplings, lower resolution is sufficient.
\begin{table}[htbp]
\caption{\label{tab:FSprobmax4}
Fock sector probabilities
$\int|\phi_{\sigma s}^{(n_i)}|^2\prod_j d\underline{q}_j$, 
where $(n_i)\equiv\{n_0,n_1,n_2,n_3\}$, 
$n=n_0$ is the number of physical bosons and $n_i$ the number of
Pauli--Villars bosons of type $i$.
The helicities $\sigma$ and $s$ refer to the
physical and bare fermion, respectively.
The numerical and physical parameters are $K=17$,
$N_\perp=5$, $M^2=\mu^2$, $\Lambda^2=50\mu^2$,
$\mu_1^2=10\mu^2$, $\mu_2^2=20\mu^2$, $\mu_3^2=30\mu^2$, and
$\langle:\!\!\phi^2(0)\!\!:\rangle=0.5$. 
Probabilities smaller than $\sim10^{-5}$ are
not resolved with any accuracy.} 
\begin{tabular}{ccccc|ccc|r}
\multicolumn{5}{c|}{Number of bosons} & 
\multicolumn{3}{c|}{Probability} &  \# of \\
\cline{1-8} 
total & $n$ & $n_1$ & $n_2$ & $n_3$ & $\sigma=s$  & $\sigma=-s$ 
                                          & total &  states \\
\hline
0 & 0 & 0 & 0 & 0 & 0.9461  & --- & 0.9461 &       1 \\
1 & 1 & 0 & 0 & 0 & 0.0593 & 0.0444 & 0.1037  &     332 \\
1 & 0 & 1 & 0 & 0 & 0.0340 & 0.0571 & 0.0911 &     219 \\
1 & 0 & 0 & 1 & 0 & 0.0182 & 0.0300 & 0.0481 &     117 \\
1 & 0 & 0 & 0 & 1 & 0.0036 & 0.0044 & 0.0080 &      55 \\
2 & 2 & 0 & 0 & 0 & 0.0014 & 0.0021 & 0.0036 &   12414 \\
2 & 1 & 1 & 0 & 0 & 0.0021 & 0.0022 & 0.0043 &    9969 \\
2 & 0 & 2 & 0 & 0 & 0.0006 & 0.0003 & 0.0009 &    1499 \\
2 & 1 & 0 & 1 & 0 & 0.0007 & 0.0007 & 0.0014 &    2998 \\
2 & 0 & 1 & 1 & 0 & 0.0002 & 0.0001 & 0.0003 &     598 \\
2 & 0 & 0 & 2 & 0 & 0.5$\cdot 10^{-5}$ & 0.4$\cdot 10^{-5}$ 
                             & 0.9$\cdot 10^{-5}$    &      25 \\
2 & 1 & 0 & 0 & 1 & 0.7$\cdot 10^{-4}$ & 0.7$\cdot 10^{-4}$
                             & 0.0001   &     655 \\
2 & 0 & 1 & 0 & 1 & 0.7$\cdot 10^{-5}$ & 0.6$\cdot 10^{-5}$ 
                             & 0.1$\cdot 10^{-4}$    &      45 \\
3 & 3 & 0 & 0 & 0 & 0.0001 & 0.2$\cdot 10^{-4}$ & 0.0001  &  136568 \\
3 & 2 & 1 & 0 & 0 & 0.0002 & 0.3$\cdot 10^{-4}$ & 0.0002  &  102490 \\
3 & 1 & 2 & 0 & 0 & 0.6$\cdot 10^{-4}$ & 0.1$\cdot 10^{-4}$ 
                             & 0.7$\cdot 10^{-4}$ &   18021 \\
3 & 0 & 3 & 0 & 0 & 0.4$\cdot 10^{-5}$ & 0.1$\cdot 10^{-5}$ 
                             & 0.5$\cdot 10^{-5}$ &     748 \\
3 & 2 & 0 & 1 & 0 & 0.4$\cdot 10^{-4}$ & 0.6$\cdot 10^{-5}$ 
                             & 0.5$\cdot 10^{-4}$ &   16631 \\
3 & 1 & 1 & 1 & 0 & 0.1$\cdot 10^{-4}$ & 0.3$\cdot 10^{-5}$ 
                             & 0.2$\cdot 10^{-4}$ &    2992 \\
3 & 0 & 2 & 1 & 0 & $\sim 10^{-6}$ & $\sim 10^{-6}$ & $\sim 10^{-6}$ &     79 \\
4 & 4 & 0 & 0 & 0 & $\sim 10^{-6}$ & $\sim 10^{-6}$ & $\sim 10^{-6}$ & 624372 \\
4 & 3 & 1 & 0 & 0 & $\sim 10^{-6}$ & $\sim 10^{-6}$ & $\sim 10^{-6}$ & 381016 \\
4 & 2 & 2 & 0 & 0 & $\sim 10^{-6}$ & $\sim 10^{-6}$ & $\sim 10^{-6}$ &  57132 \\
4 & 1 & 3 & 0 & 0 & $\sim 10^{-7}$ & $\sim 10^{-7}$ & $\sim 10^{-7}$ &   2577 \\
4 & 0 & 4 & 0 & 0 & $\sim 10^{-9}$ & $\sim 10^{-9}$ & $\sim 10^{-9}$ &     30 \\
4 & 3 & 0 & 1 & 0 & $\sim 10^{-7}$ & $\sim 10^{-6}$ & $\sim 10^{-6}$ &  28613 \\
4 & 2 & 1 & 1 & 0 & $\sim 10^{-7}$ & $\sim 10^{-7}$ & $\sim 10^{-7}$ &   2647 \\
\end{tabular}
\end{table}

Given these considerations for resolution and truncation, we have
done two sets of calculations with $K$ between 11 and 29, or even 39.
For $K=11$ and 13 there is no explicit truncation and the maximum
number of bosons is 5 and 6, respectively.  For $K=15$, 17, and 19,
the maximum number of bosons used is 4, and for $K \geq 21$ the maximum
used is 2.  Two different sets of cutoff and PV masses were
considered: $\Lambda^2=50\mu^2$, $\mu_1^2=10\mu^2$, $\mu_2^2=20\mu^2$,
and $\mu_3^2=30\mu^2$; and $\Lambda^2=100\mu^2$, $\mu_1^2=20\mu^2$, 
$\mu_2^2=40\mu^2$, and $\mu_3^2=60\mu^2$.  The transverse resolution
$N_\perp$ was at least 4 and was increased beyond that to the extent
allowed by the available storage on a 16 GB node of an IBM SP.  The
four processors of the node were used in parallel.

A sampling of explored parameter values can be seen in
Tables~\ref{tab:obs010L50-1}-\ref{tab:obs100L100-2}\@.  For each choice
of input parameter values, these tables present the results
for the bare parameters of the Hamiltonian, $g$ and $\delta M^2$,
and for various expectation values.
\begin{table}[htbp]
\squeezetable
\caption{\label{tab:obs010L50-1}
Bare parameters and observables.
The input parameter values are $M^2=\mu^2$, $\Lambda^2=50\mu^2$,
$\mu_1^2=10\mu^2$, $\mu_2^2=20\mu^2$, and $\mu_3^2=30\mu^2$.}
\begin{center}
\begin{tabular}{cc|c|cc|ccccc}
 $K$ &  $N_\perp$ & $\langle:\!\!\phi^2(0)\!\!:\rangle$ &  $g$   
    &   $\delta M^2/\mu^2$ & $|\psi_0|^2$ & $-100\mu^2F'(0)$  
    & $\frac{(\langle y_1 y_2\rangle-\langle y\rangle^2)_{n\geq 2}}
            {\langle y\rangle^2}$  \\
\hline
11 &  4 &  0.100 &  1.279 &  0.062 &  0.988 &  0.090 & 3653. \\
11 &  5 &  0.100 &  1.339 &  0.060 &  0.989 &  0.092 & 3938. \\
11 &  6 &  0.100 &  1.366 &  0.066 &  0.989 &  0.095 & 3150. \\
13 &  4 &  0.100 &  1.274 &  0.035 &  0.988 &  0.092 & 4117. \\
13 &  5 &  0.100 &  1.335 &  0.037 &  0.989 &  0.093 & 3426. \\
15 &  4 &  0.100 &  1.306 &  0.084 &  0.988 &  0.086 & 3671. \\
15 &  5 &  0.100 &  1.361 &  0.087 &  0.988 &  0.091 & 3108. \\
17 &  4 &  0.100 &  1.292 &  0.055 &  0.988 &  0.088 & 3331. \\
17 &  5 &  0.100 &  1.342 &  0.055 &  0.989 &  0.097 & 2868. \\
19 &  4 &  0.100 &  1.287 &  0.054 &  0.988 &  0.091 & 3324. \\
19 &  5 &  0.100 &  1.349 &  0.056 &  0.988 &  0.096 & 3375. \\
21 & 4 & 0.100 &  1.281 &  0.055 &  0.988 &  0.088 &  4075. \\
21 & 5 & 0.100 &  1.350 &  0.057 &  0.988 &  0.095 &  3257.\\
21 & 6 & 0.100 &  1.378 &  0.058 &  0.989 &  0.099 &  2788.\\
21 & 7 & 0.100 &  1.392 &  0.058 &  0.989 &  0.101 &  2977.\\
21 & 8 & 0.100 &  1.400 &  0.059 &  0.989 &  0.104 &  2920. \\
29 & 4 & 0.100 &  1.302 &  0.062 &  0.987 &  0.091 &  3572. \\
29 & 5 & 0.100 &  1.353 &  0.063 &  0.988 &  0.096 &  3441. \\
29 & 6 & 0.100 &  1.386 &  0.065 &  0.989 &  0.099 &  2920. \\
29 & 7 & 0.100 &  1.400 &  0.066 &  0.989 &  0.103 &  2983. \\
\end{tabular}
\end{center}
\end{table}
\begin{table}[htbp]
\squeezetable
\caption{\label{tab:obs010L50-2}
Additional observables.
The input parameter values are the same as for
Table~\ref{tab:obs010L50-1}.
The dashes represent values not extracted from preliminary,
lower-resolution calculations.}
\begin{center}
\begin{tabular}{cc|c|ccc|ccc}
 $K$ &  $N_\perp$ & $\langle:\!\!\phi^2(0)\!\!:\rangle$    
    & $\langle n_{B,\sigma}\rangle$ & $\langle n_{B,-\sigma}\rangle$ & 
                      $\langle n_B\rangle$ 
    & $\langle y_{B,\sigma}\rangle$ & $\langle y_{B,-\sigma}\rangle$ & 
                      $\langle y_B\rangle$  \\
\hline
11 & 4 & 0.100 & -- &  -- &  0.021 & -- &  -- &  0.01191  \\
11 & 5 & 0.100 & -- &  -- &  0.021 & -- &  -- &  0.01209  \\
11 & 6 & 0.100 & -- &  -- &  0.021 & -- &  -- &  0.01213  \\
13 & 4 & 0.100 & -- &  -- &  0.021 & -- &  -- &  0.01170  \\
13 & 5 & 0.100 & -- &  -- &  0.022 & -- &  -- &  0.01219  \\
15 & 4 & 0.100 & -- &  -- &  0.021 & -- &  -- &  0.01198  \\
15 & 5 & 0.100 & -- &  -- &  0.022 & -- &  -- &  0.01225  \\
17 & 4 & 0.100 & -- &  -- &  0.022 & -- &  -- &  0.01220  \\
17 & 5 & 0.100 & -- &  -- &  0.022 & -- &  -- &  0.01234  \\
19 & 4 & 0.100 & -- &  -- &  0.022 & -- &  -- &  0.01231  \\
19 & 5 & 0.100 & -- &  -- &  0.022 & -- &  -- &  0.01238  \\
21 & 4 & 0.100 &  0.0147 &  0.0072 &  0.0219 &  0.00836 &  0.00417 &  0.01253 \\
21 & 5 & 0.100 &  0.0133 &  0.0086 &  0.0218 &  0.00735 &  0.00506 &  0.01241 \\
21 & 6 & 0.100 &  0.0128 &  0.0091 &  0.0220 &  0.00712 &  0.00540 &  0.01252 \\
21 & 7 & 0.100 &  0.0125 &  0.0094 &  0.0219 &  0.00695 &  0.00556 &  0.01251 \\
21 & 8 & 0.100 &  0.0125 &  0.0096 &  0.0220 &  0.00691 &  0.00566 &  0.01257 \\
29 & 4 & 0.100 &  0.0145 &  0.0076 &  0.0221 &  0.00808 &  0.00444 &  0.01252 \\
29 & 5 & 0.100 &  0.0135 &  0.0086 &  0.0221 &  0.00753 &  0.00508 &  0.01260 \\
29 & 6 & 0.100 &  0.0129 &  0.0092 &  0.0221 &  0.00711 &  0.00546 &  0.01257 \\
29 & 7 & 0.100 &  0.0126 &  0.0095 &  0.0221 &  0.00697 &  0.00564 &  0.01261 \\
\end{tabular}
\end{center}
\end{table}
\begin{table}[htbp]
\squeezetable
\caption{\label{tab:obs100L50-1}
Same as Table~\ref{tab:obs010L50-1}, but for stronger coupling.}
\begin{center}
\begin{tabular}{cc|c|cc|ccccc}
 $K$ &  $N_\perp$ & $\langle:\!\!\phi^2(0)\!\!:\rangle$ &  $g$   
    &   $\delta M^2/\mu^2$ & $|\psi_0|^2$ & $-100\mu^2F'(0)$  
    & $\frac{(\langle y_1 y_2\rangle-\langle y\rangle^2)_{n\geq 2}}
            {\langle y\rangle^2}$  \\
\hline
11 &  4 &  1.000 &  5.417 &  0.863 &  0.891 &  1.340 &  21.42 \\
11 &  5 &  1.001 &  5.325 &  0.699 &  0.899 &  1.073 &  27.77 \\
11 &  6 &  1.000 &  5.105 &  0.819 &  0.883 &  2.849 &  22.06 \\
13 &  4 &  1.000 &  4.520 &  0.362 &  0.886 &  0.929 &  36.62 \\
13 &  5 &  0.998 &  4.746 &  0.362 &  0.893 &  1.244 &  28.91 \\
15 &  4 &  1.000 &  5.783 &  1.161 &  0.918 &  0.457 &  22.85 \\
15 &  5 &  1.000 &  5.589 &  1.124 &  0.913 &  0.641 &  23.52 \\
17 &  4 &  1.000 &  5.045 &  0.619 &  0.909 &  0.683 &  25.69 \\
17 &  5 &  0.998 &  4.806 &  0.562 &  0.899 &  1.032 &  26.31 \\
19 &  4 &  1.000 &  4.804 &  0.576 &  0.895 &  0.853 &  28.82 \\
19 &  5 &  1.000 &  4.973 &  0.597 &  0.902 &  0.868 &  29.83 \\
21 & 4 & 1.000 &  4.925 &  0.615 &  0.896 &  0.774 &    32.23 \\
21 & 5 & 1.000 &  5.082 &  0.616 &  0.903 &  0.876 &    28.27 \\
21 & 6 & 1.000 &  4.978 &  0.619 &  0.900 &  1.266 &    25.64 \\
21 & 7 & 0.999 &  5.046 &  0.612 &  0.903 &  1.140 &    27.78 \\
21 & 8 & 1.000 &  5.009 &  0.622 &  0.901 &  1.248 &    27.68 \\
29 & 4 & 1.000 &  5.089 &  0.713 &  0.902 &  0.659 &    29.23 \\
29 & 5 & 1.000 &  5.072 &  0.695 &  0.901 &  0.837 &    29.11 \\
29 & 6 & 1.000 &  5.173 &  0.708 &  0.907 &  0.867 &    26.29 \\
29 & 7 & 0.999 &  5.152 &  0.712 &  0.905 &  0.933 &    27.62 \\
\end{tabular}
\end{center}
\end{table}
\begin{table}[htbp]
\squeezetable
\caption{\label{tab:obs100L50-2}
Same as Table~\ref{tab:obs010L50-2}, but for stronger coupling.}
\begin{center}
\begin{tabular}{cc|c|ccc|ccc}
 $K$ &  $N_\perp$ & $\langle:\!\!\phi^2(0)\!\!:\rangle$    
    & $\langle n_{B,\sigma}\rangle$ & $\langle n_{B,-\sigma}\rangle$ & 
                           $\langle n_B\rangle$ 
    & $\langle y_{B,\sigma}\rangle$ & $\langle y_{B,-\sigma}\rangle$ & 
                              $\langle y_B\rangle$  \\
\hline
11 & 4 & 1.000 & -- &  -- &  0.237 & -- &  -- & 0.1425 \\
11 & 5 & 1.001 & -- &  -- &  0.231 & -- &  -- & 0.1342 \\
11 & 6 & 1.000 & -- &  -- &  0.233 & -- &  -- & 0.1383 \\
13 & 4 & 1.000 & -- &  -- &  0.217 & -- &  -- & 0.1205 \\
13 & 5 & 0.998 & -- &  -- &  0.222 & -- &  -- & 0.1280 \\
15 & 4 & 1.000 & -- &  -- &  0.210 & -- &  -- & 0.1167 \\
15 & 5 & 1.000 & -- &  -- &  0.212 & -- &  -- & 0.1181 \\
17 & 4 & 1.000 & -- &  -- &  0.211 & -- &  -- & 0.1192 \\
17 & 5 & 0.998 & -- &  -- &  0.214 & -- &  -- & 0.1212 \\
19 & 4 & 1.000 & -- &  -- &  0.216 & -- &  -- & 0.1220 \\
19 & 5 & 1.000 & -- &  -- &  0.216 & -- &  -- & 0.1217 \\
21 & 4 & 1.000 &  0.1179 &  0.0994 &  0.2173 &  0.06625 &  0.05649 &  0.12275 \\
21 & 5 & 1.000 &  0.1044 &  0.1129 &  0.2174 &  0.05788 &  0.06580 &  0.12368 \\
21 & 6 & 1.000 &  0.1031 &  0.1154 &  0.2185 &  0.05770 &  0.06759 &  0.12529 \\
21 & 7 & 1.000 &  0.1008 &  0.1162 &  0.2170 &  0.05575 &  0.06751 &  0.12326 \\
21 & 8 & 1.000 &  0.1016 &  0.1153 &  0.2169 &  0.05608 &  0.06655 &  0.12263 \\
29 & 4 & 1.000 &  0.1088 &  0.1057 &  0.2145 &  0.05894 &  0.06072 &  0.11966 \\
29 & 5 & 1.000 &  0.1048 &  0.1122 &  0.2170 &  0.05738 &  0.06452 &  0.12190 \\
29 & 6 & 1.000 &  0.0982 &  0.1178 &  0.2160 &  0.05363 &  0.06824 &  0.12187 \\
29 & 7 & 1.000 &  0.0985 &  0.1180 &  0.2165 &  0.05360 &  0.06803 &  0.12163 \\
%\hline
\end{tabular}
\end{center}
\end{table}
\begin{table}[htbp]
\squeezetable
\caption{\label{tab:obs100L100-1}
Same as Table~\ref{tab:obs100L50-1}, but for $\Lambda^2=100\mu^2$,
$\mu_1^2=20\mu^2$, $\mu_2^2=40\mu^2$, and $\mu_3^2=60\mu^2$.}
\begin{center}
\begin{tabular}{cc|c|cc|ccccc}
 $K$ &  $N_\perp$ & $\langle:\!\!\phi^2(0)\!\!:\rangle$ &  $g$   
    &   $\delta M^2/\mu^2$ & $|\psi_0|^2$ & $-100\mu^2F'(0)$  
    & $\frac{(\langle y_1 y_2\rangle-\langle y\rangle^2)_{n\geq 2}}
            {\langle y\rangle^2}$  \\
\hline
11 & 4 & 0.989 & 4.661 & 1.220 & 0.858 & 1.195 & 22.07 \\
11 & 5 & 0.998 & 4.326 & 0.991 & 0.868 & 1.114 & 26.07 \\
11 & 6 & 0.904 & 4.181 & 1.063 & 0.882 & 1.442 & 28.64 \\
13 & 4 & 1.002 & 3.378 & 0.487 & 0.836 & 1.122 & 32.87 \\
13 & 5 & 1.000 & 4.182 & 0.537 & 0.863 & 1.245 & 27.45 \\
15 & 4 & 1.000 & 5.332 & 1.849 & 0.912 & 0.337 & 19.69 \\
15 & 5 & 1.000 & 5.142 & 1.804 & 0.908 & 0.458 & 21.64 \\
17 & 4 & 0.998 & 4.448 & 1.131 & 0.896 & 0.448 & 28.27 \\
17 & 5 & 1.000 & 4.527 & 1.094 & 0.889 & 0.758 & 24.09 \\
19 & 4 & 1.000 & 4.317 & 0.915 & 0.873 & 0.759 & 27.20 \\
19 & 5 & 0.999 & 4.412 & 0.970 & 0.880 & 0.960 & 24.52 \\
21 & 4 & 1.000 & 4.300 & 0.966 & 0.876 & 0.752 & 25.75 \\
29 & 4 & 1.000 & 4.795 &  1.160 &  0.886 &  0.565 &    23.4 \\
29 & 5 & 0.999 & 4.683 &  1.132 &  0.893 &  0.699 &    22.5 \\
29 & 6 & 1.000 & 4.772 &  1.164 &  0.897 &  0.757 &    21.9 \\
29 & 7 & 0.999 & 4.706 &  1.160 &  0.893 &  1.079 &    20.9 \\
39 & 4 & 0.998 & 4.562 &  1.006 &  0.884 &  0.660 &    23.6 \\
39 & 5 & 1.000 & 4.678 &  1.026 &  0.892 &  0.729 &    22.3 \\
39 & 6 & 1.000 & 4.656 &  1.020 &  0.894 &  0.833 &    21.8 \\
\end{tabular}
\end{center}
\end{table}
\begin{table}[htbp]
\squeezetable
\caption{\label{tab:obs100L100-2}
Same as Table~\ref{tab:obs100L50-2}, but for $\Lambda^2=100\mu^2$,
$\mu_1^2=20\mu^2$, $\mu_2^2=40\mu^2$, and $\mu_3^2=60\mu^2$.}
\begin{center}
\begin{tabular}{cc|c|ccc|ccc}
 $K$ &  $N_\perp$ & $\langle:\!\!\phi^2(0)\!\!:\rangle$    
    & $\langle n_{B,\sigma}\rangle$ & $\langle n_{B,-\sigma}\rangle$ & 
                             $\langle n_B\rangle$ 
    & $\langle y_{B,\sigma}\rangle$ & $\langle y_{B,-\sigma}\rangle$ & 
                             $\langle y_B\rangle$  \\
\hline
11 & 4 & 0.989 & -- &  -- &   0.245 & -- &  -- &  0.1524 \\
11 & 5 & 0.998 & -- &  -- &   0.227 & -- &  -- &   0.1310 \\
11 & 6 & 0.904 & -- &  -- &   0.206 & -- &  -- &   0.1207 \\
13 & 4 & 1.002 & -- &  -- &   0.228 & -- &  -- &   0.1297 \\
13 & 5 & 1.000 & -- &  -- &   0.230 & -- &  -- &   0.1370 \\
15 & 4 & 1.000 & -- &  -- &  0.205 & -- &  -- &   0.1149 \\
15 & 5 & 1.000 & -- &  -- &   0.204 & -- &  -- &   0.1140 \\
17 & 4 & 0.998 & -- &  -- &   0.198 & -- &  -- &   0.1078 \\
17 & 5 & 1.000 & -- &  -- &   0.212 & -- &  -- &   0.1203 \\
19 & 4 & 1.000 & -- &  -- &   0.218 & -- &  -- &   0.1261 \\
19 & 5 & 0.999 & -- &  -- &   0.218 & -- &  -- &   0.1271 \\
21 & 4 & 1.000 & -- &  -- &   0.215 & -- &  -- &   0.1245 \\
29 & 4 & 1.000 &  0.1051 &  0.1152 &  0.2203 &  0.06028 &  0.06653 &  0.12680 \\
29 & 5 & 1.000 &  0.0894 &  0.1231 &  0.2125 &  0.04925 &  0.07260 &  0.12185 \\
29 & 6 & 1.000 &  0.0798 &  0.1317 &  0.2115 &  0.04334 &  0.07794 &  0.12128 \\
29 & 7 & 1.000 &  0.0787 &  0.1350 &  0.2137 &  0.04357 &  0.08055 &  0.12412 \\
39 & 4 & 1.000 &  0.1069 &  0.1090 &  0.2159 &  0.06156 &  0.06396 &  0.12552 \\
39 & 5 & 1.000 &  0.0907 &  0.1232 &  0.2139 &  0.05109 &  0.07272 &  0.12381 \\
39 & 6 & 1.000 &  0.0836 &  0.1297 &  0.2133 &  0.04611 &  0.07694 &  0.12305 \\
\end{tabular}
\end{center}
\end{table}

\clearpage

\begin{figure}[htbp]
\begin{tabular}{cc}
\psfig{file=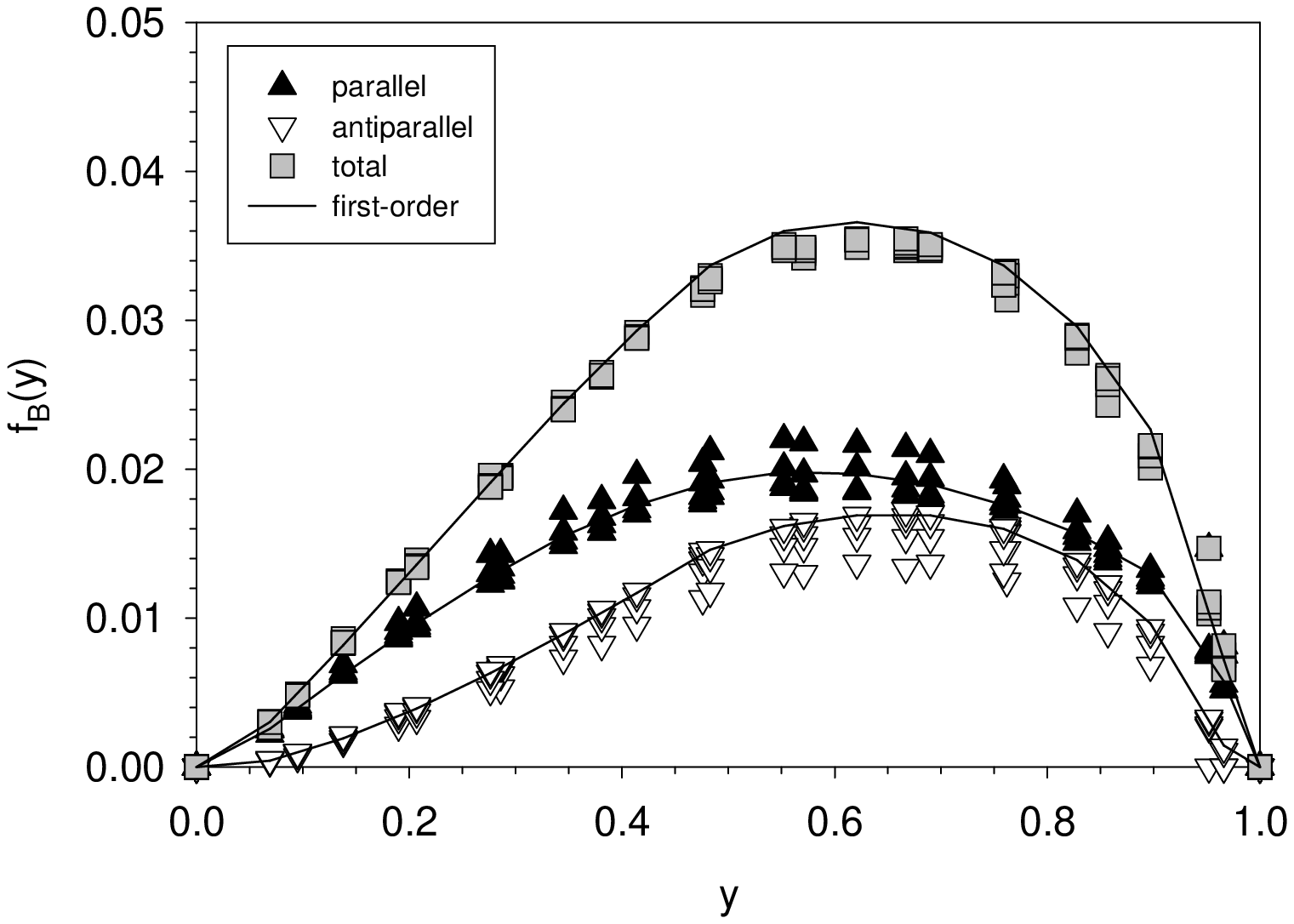,width=3.2in} &
\psfig{file=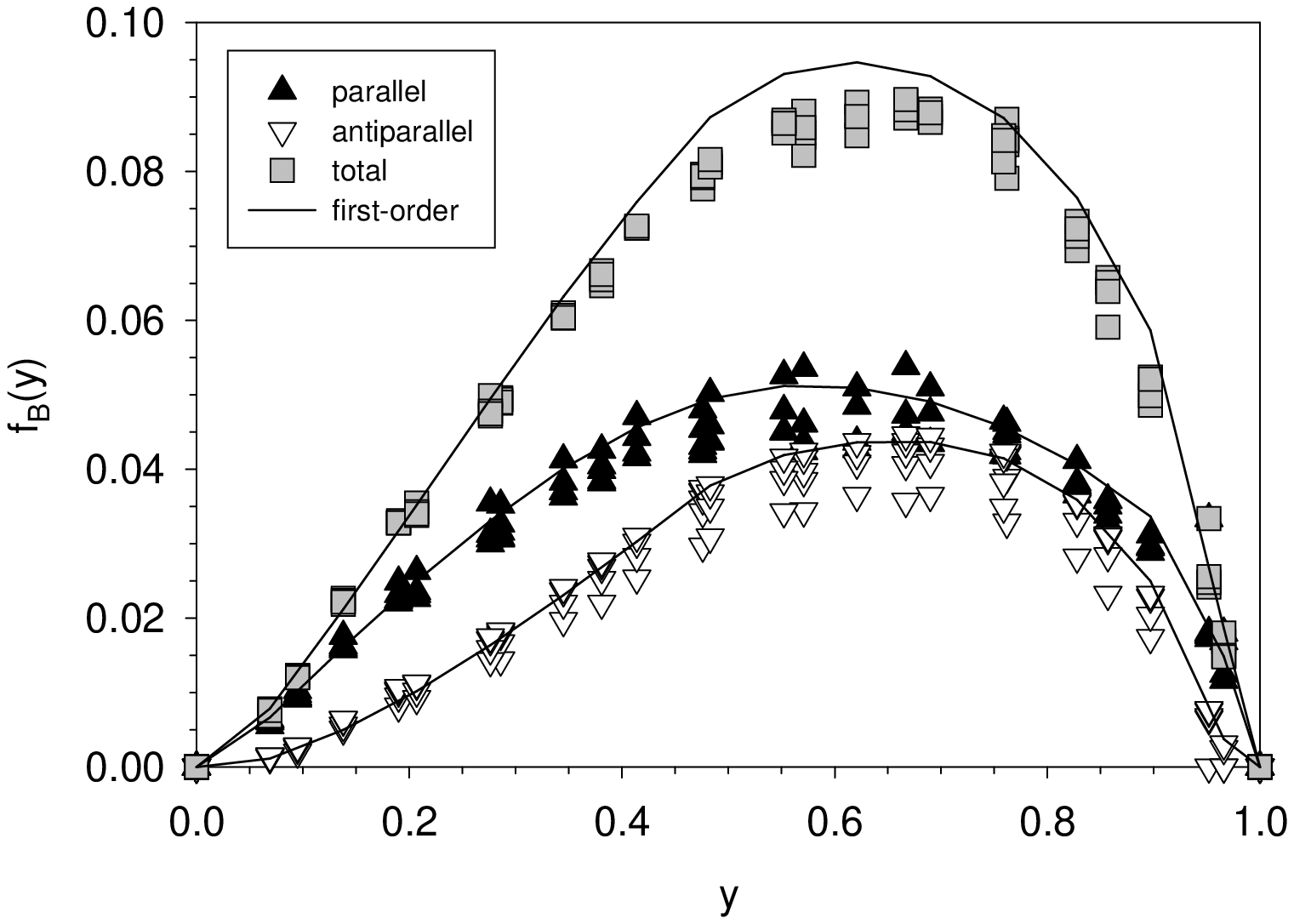,width=3.2in} \\
(a) & (b) \\
\psfig{file=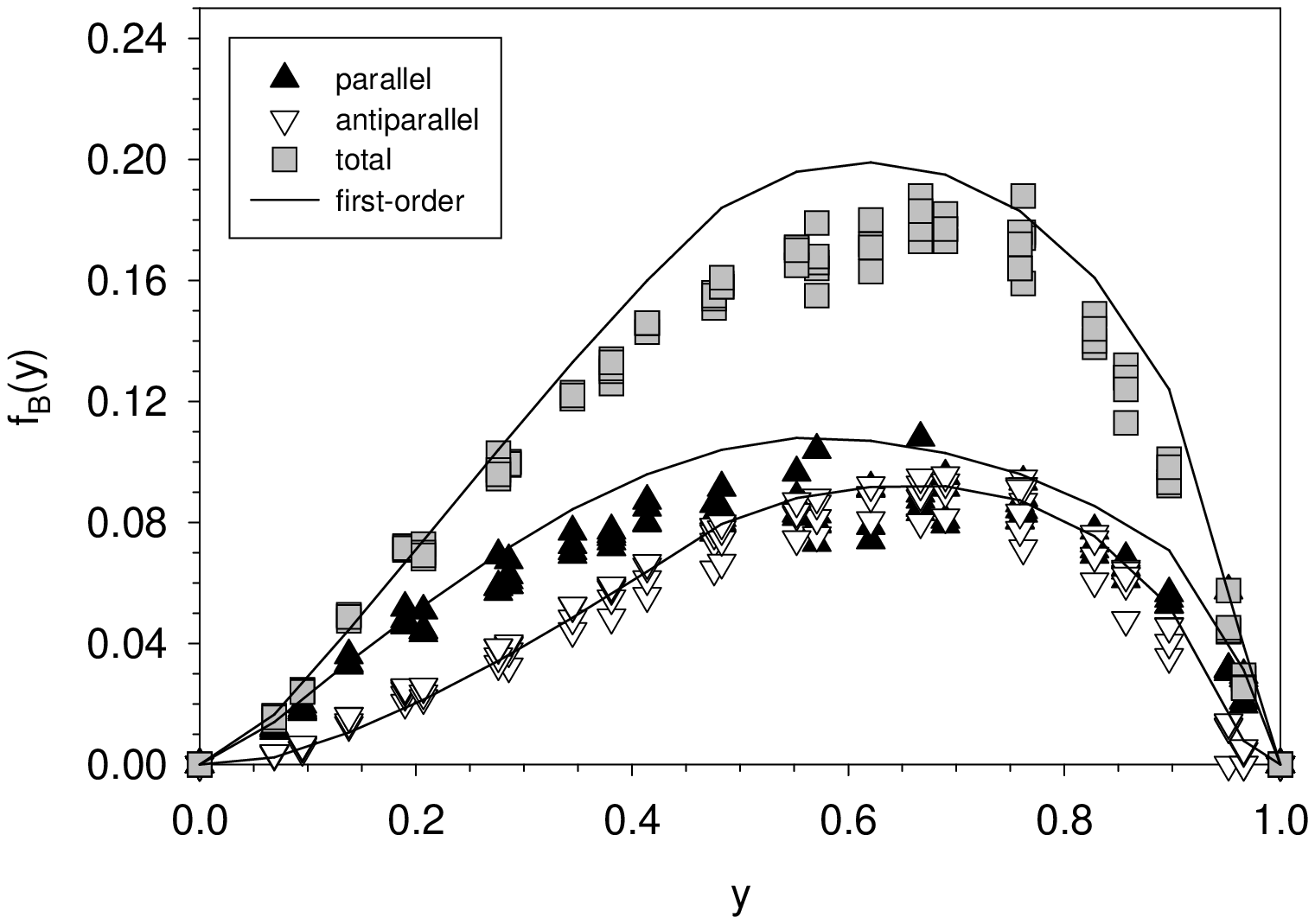,width=3.2in} &
\psfig{file=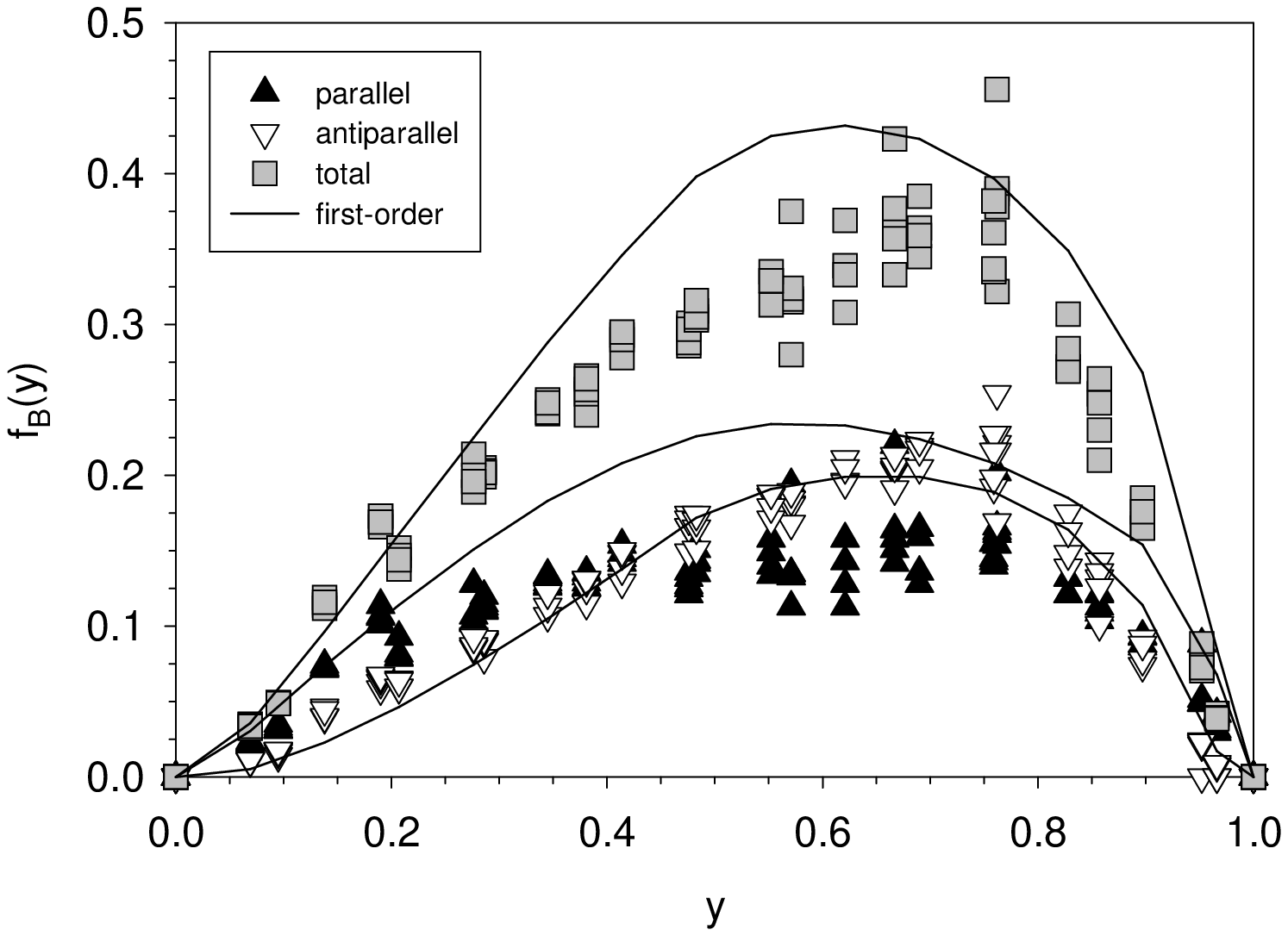,width=3.2in} \\
(c) & (d) \\
\end{tabular}
\caption{\label{fig:fBL50} The boson structure function $f_B$ at various
numerical resolutions for 
(a) $\langle:\!\!\phi^2(0)\!\!:\rangle=0.1$, 
(b) $\langle:\!\!\phi^2(0)\!\!:\rangle=0.25$, 
(c) $\langle:\!\!\phi^2(0)\!\!:\rangle=0.5$, and 
(d) $\langle:\!\!\phi^2(0)\!\!:\rangle=1.0$, 
with $M=\mu$, $\Lambda^2=50\mu^2$, $\mu_1^2=10\mu^2$, $\mu_2^2=20\mu^2$, 
and $\mu_3^2=30\mu^2$.  The solid line is from first-order perturbation
theory. } 
\end{figure}

\begin{figure}[htbp]
\begin{tabular}{cc}
\psfig{file=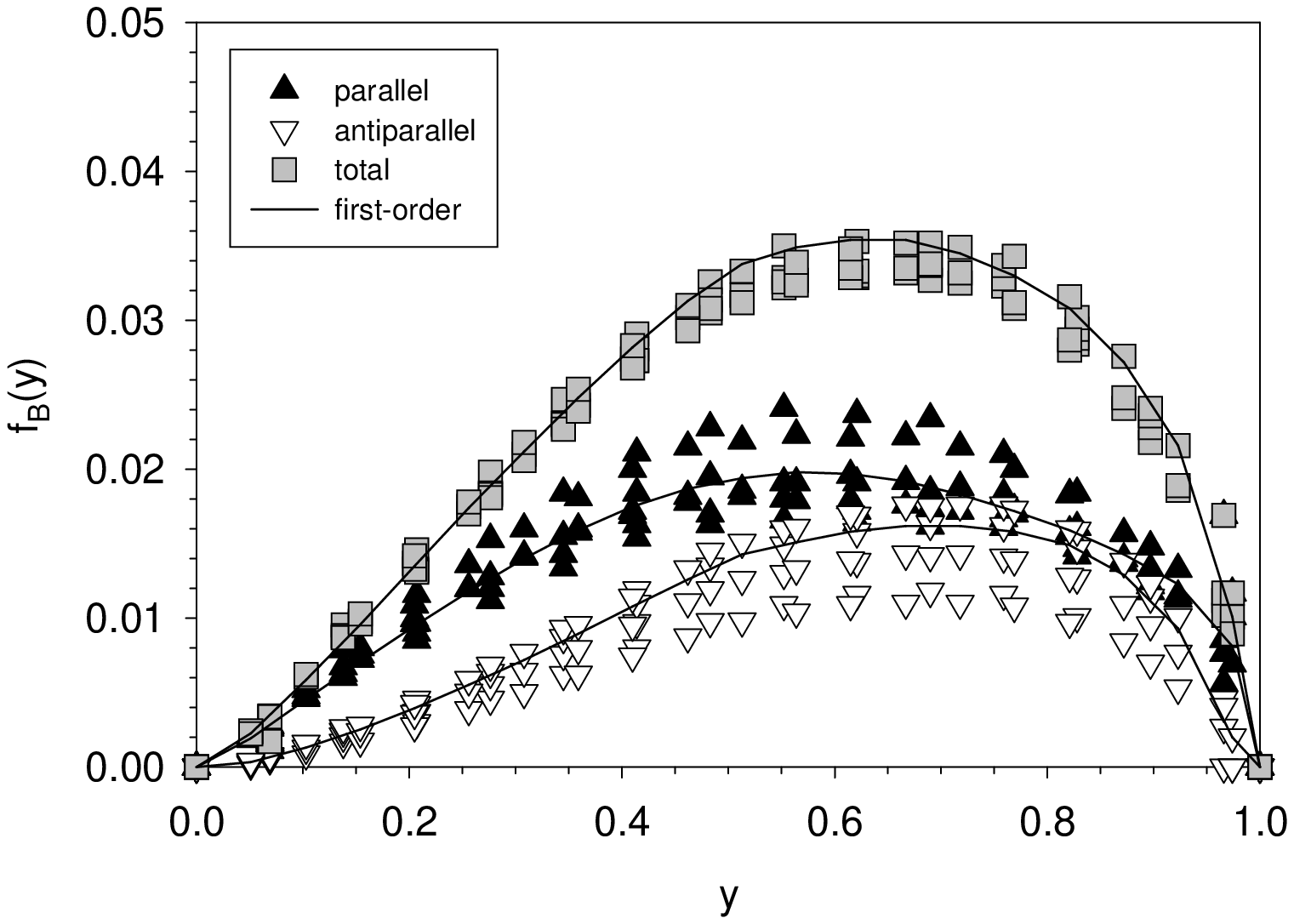,width=3.2in} &
\psfig{file=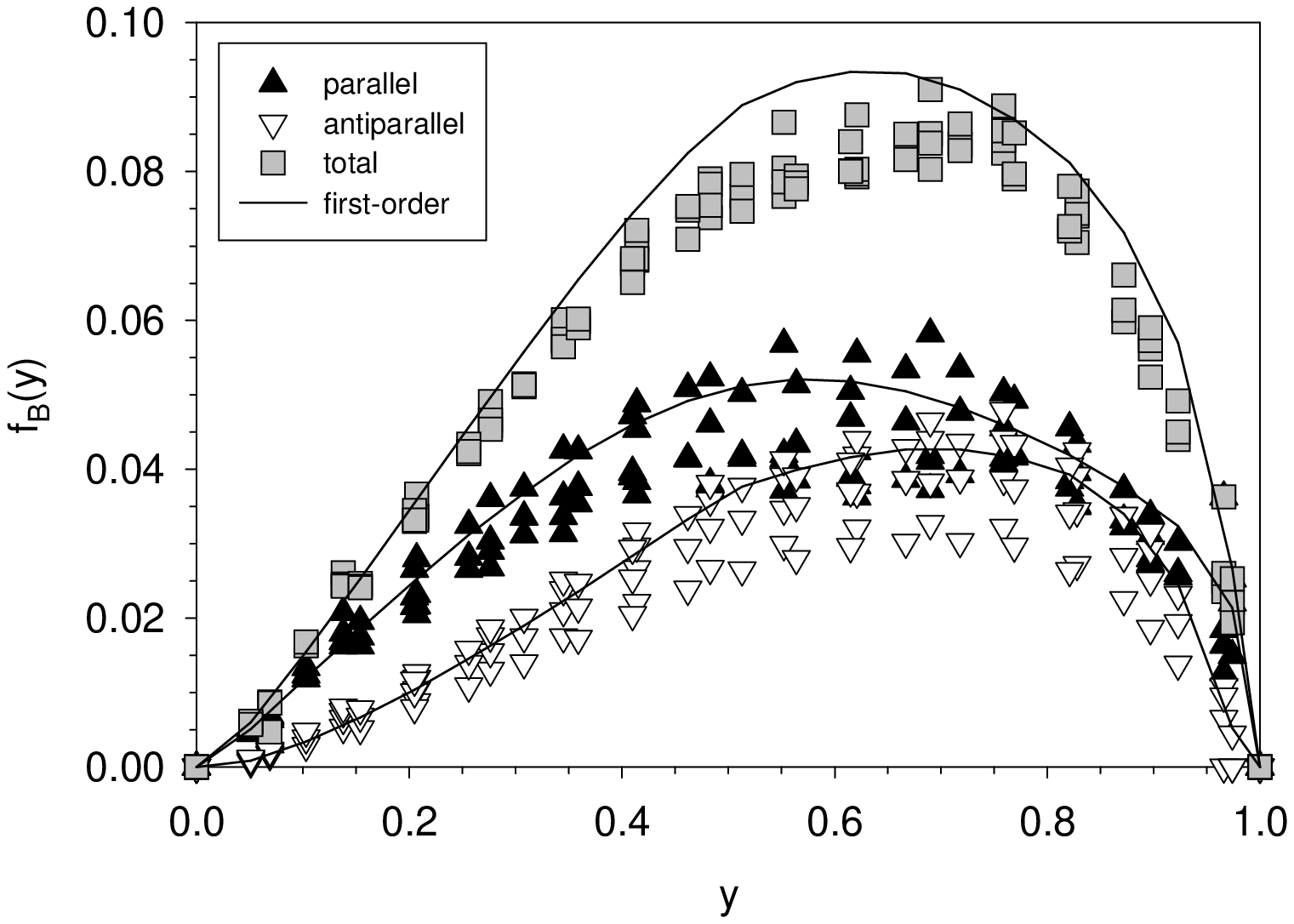,width=3.2in} \\
(a) & (b) \\
\psfig{file=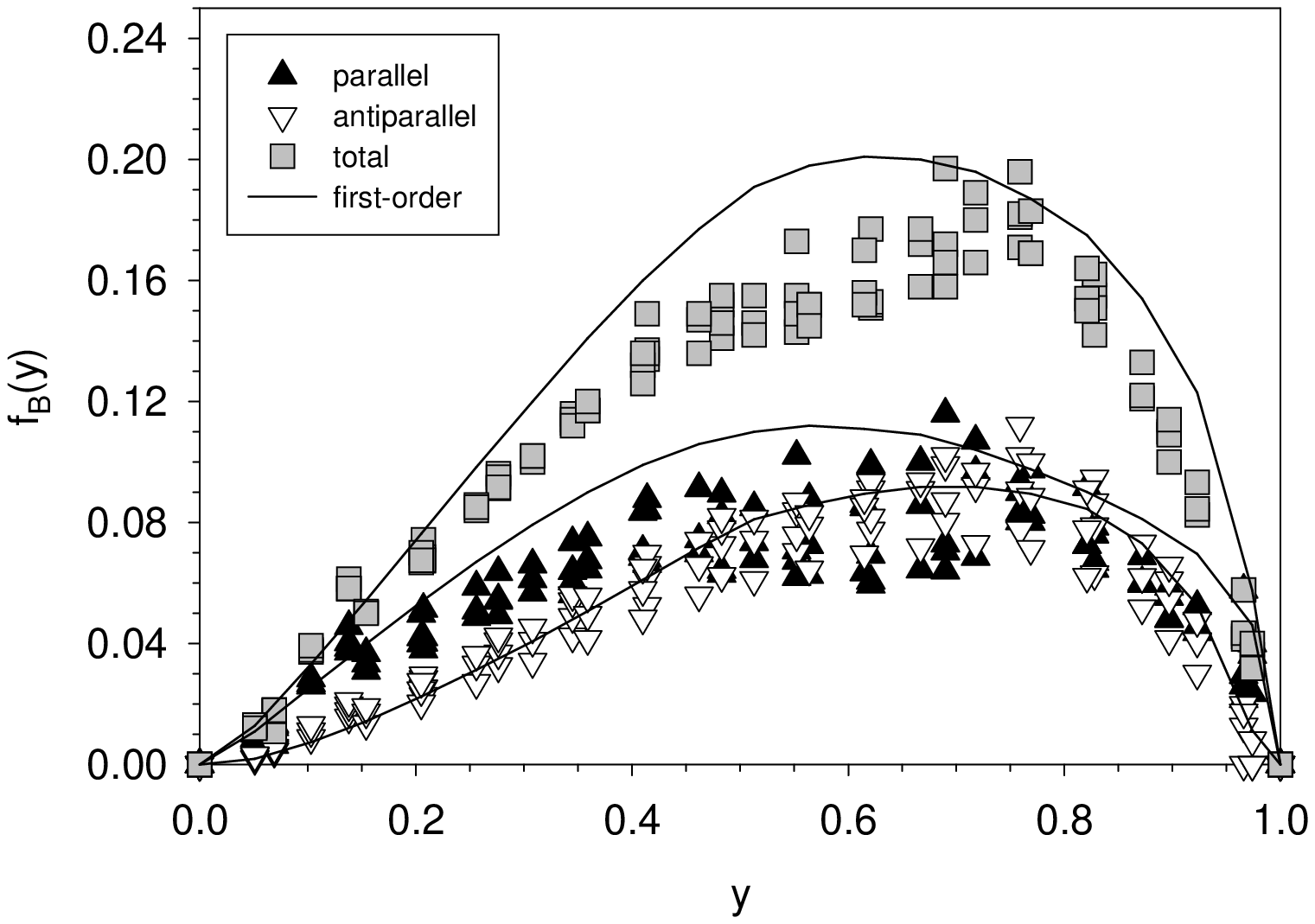,width=3.2in} &
\psfig{file=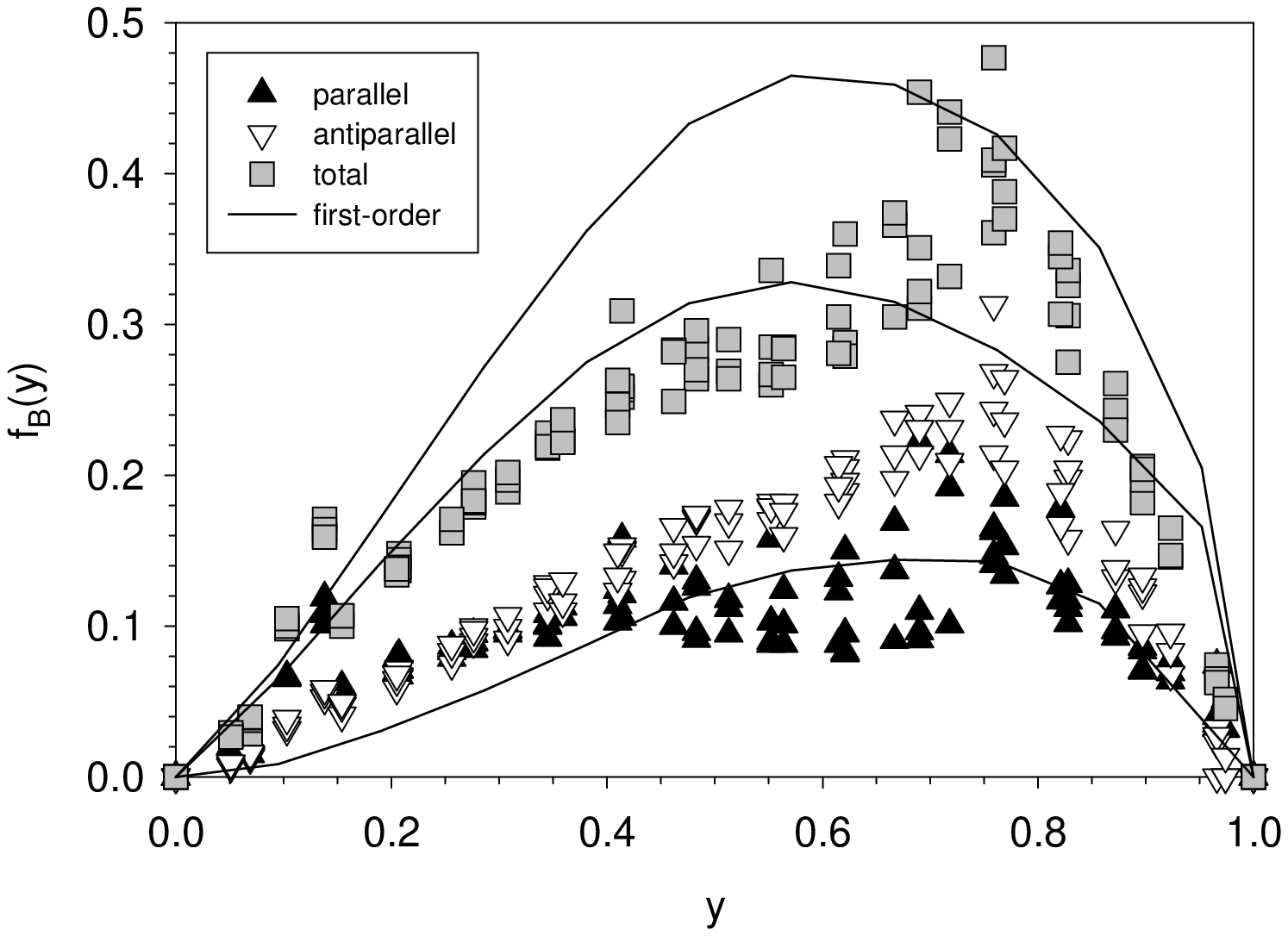,width=3.2in} \\
(c) & (d) \\
\end{tabular}
\caption{\label{fig:fBL100} 
Same as Fig.~\ref{fig:fBL50} but for
$\Lambda^2=100\mu^2$, $\mu_1^2=20\mu^2$, $\mu_2^2=40\mu^2$, and
$\mu_3^2=60\mu^2$. } 
\end{figure}

\begin{figure}[htbp]
\begin{tabular}{cc}
\psfig{file=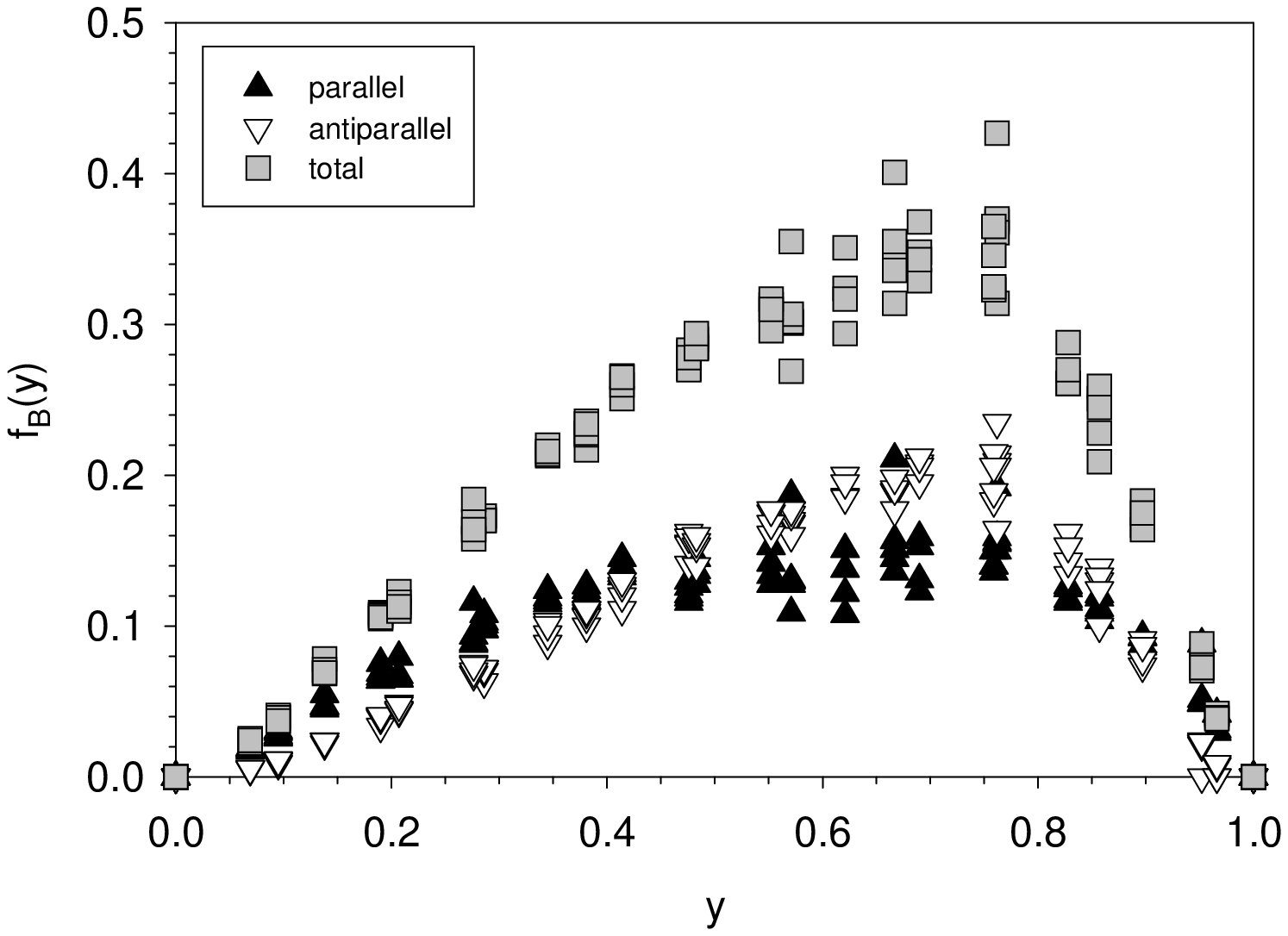,width=3.2in} &
\psfig{file=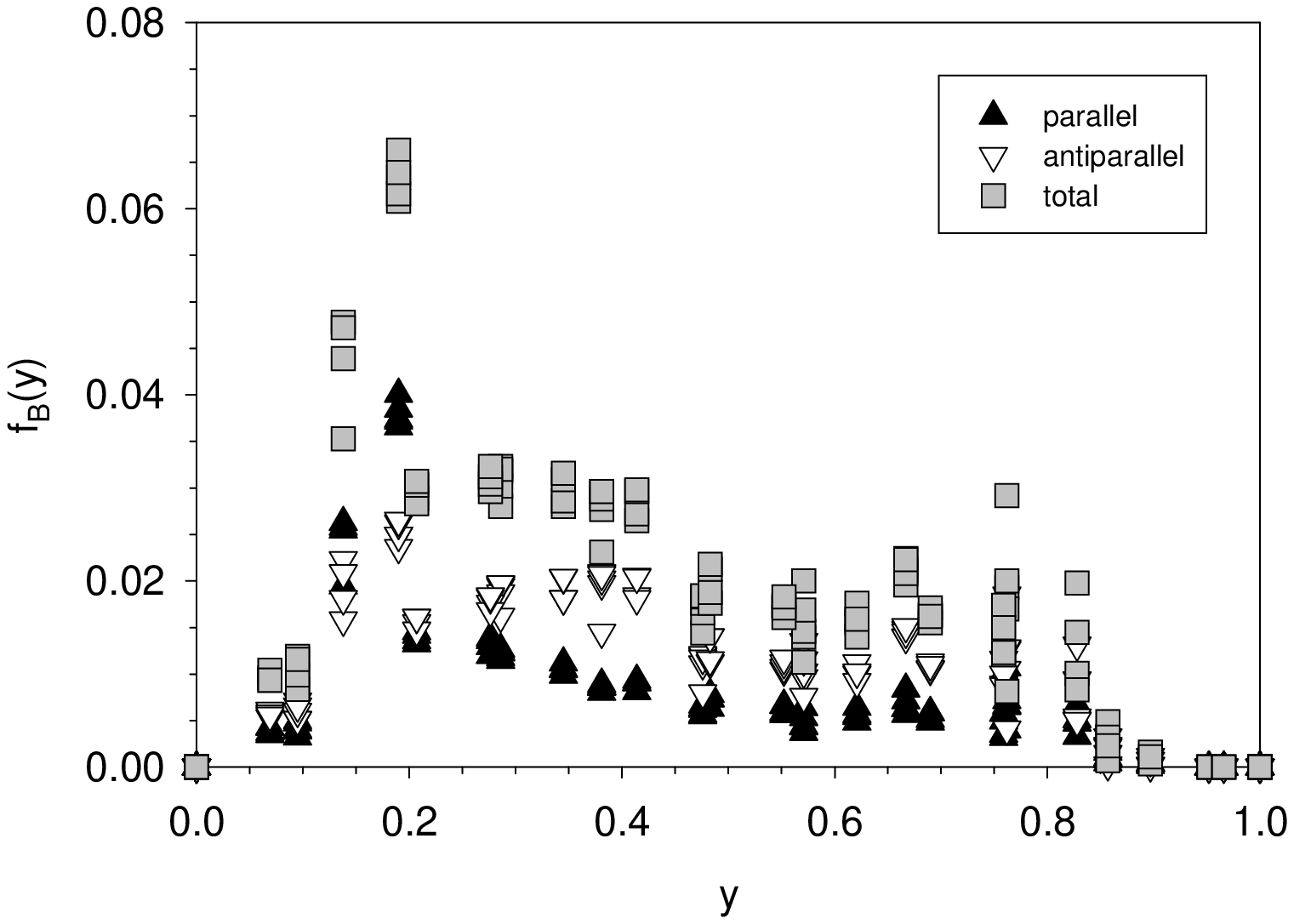,width=3.2in} \\
(a) & (b) \\
\end{tabular}
\caption{\label{fig:fB100L50onetwo} 
The (a) one-boson and (b) two-boson contributions 
to the boson structure function $f_B$ 
at various numerical resolutions, with $M=\mu$, %\protect\\
$\langle:\!\!\phi^2(0)\!\!:\rangle=1$,
$\Lambda^2=50\mu^2$, $\mu_1^2=10\mu^2$, $\mu_2^2=20\mu^2$, and
$\mu_3^2=30\mu^2$.} 
\end{figure}

\begin{figure}[htbp]
\centerline{\psfig{file=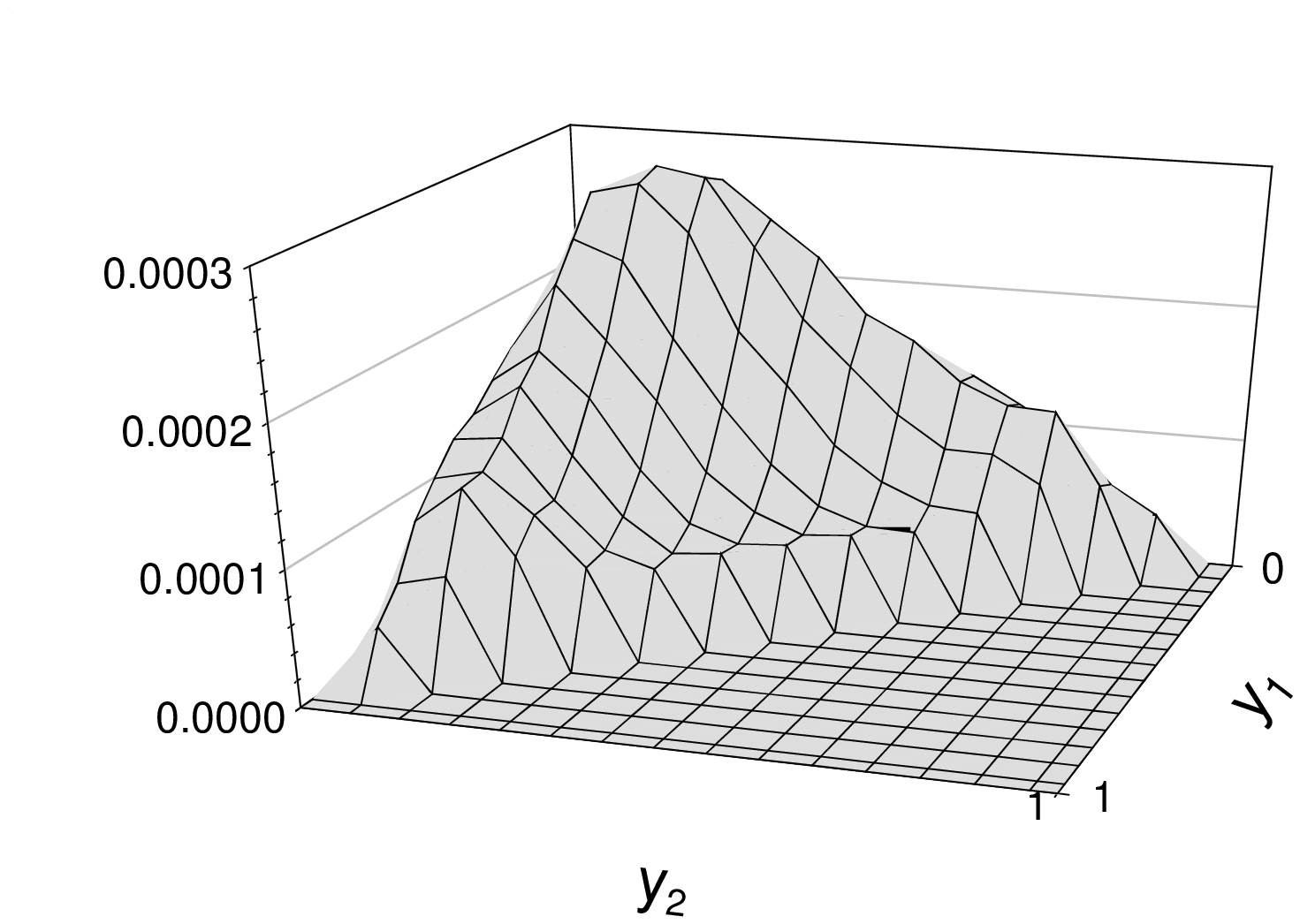,width=3.5in}}
\caption{\label{fig:fBt100L50two} 
The two-boson structure function $\tilde{f}_{B\sigma}$ 
for $K=29$ and $N_\perp=7$, with $M=\mu$, %\protect\\
$\langle:\!\!\phi^2(0)\!\!:\rangle=1$,
$\Lambda^2=50\mu^2$, $\mu_1^2=10\mu^2$, $\mu_2^2=20\mu^2$, and
$\mu_3^2=30\mu^2$. } 
\end{figure}

\begin{figure}[htbp]
\begin{tabular}{cc}
\psfig{file=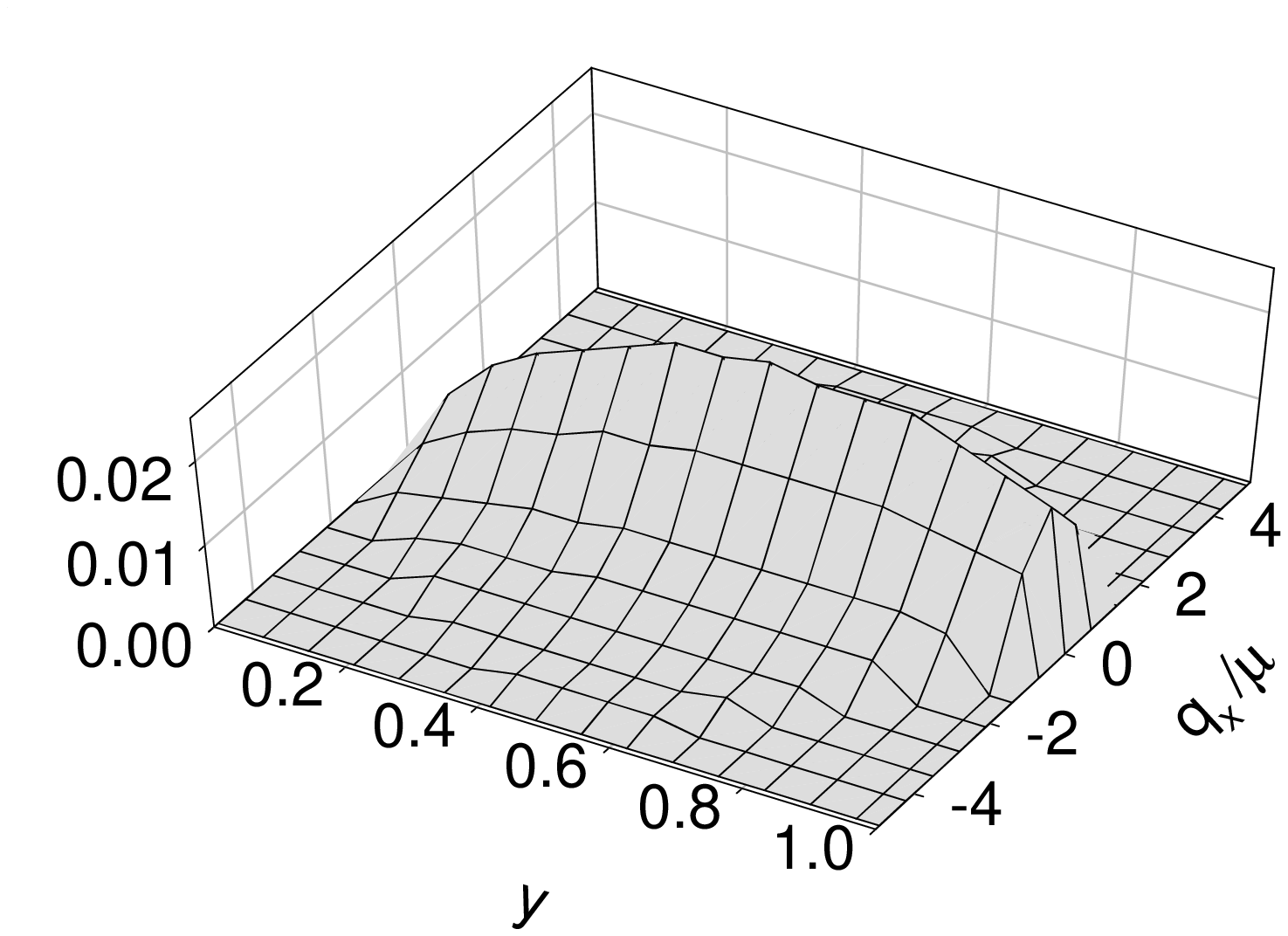,width=3.2in} &
\psfig{file=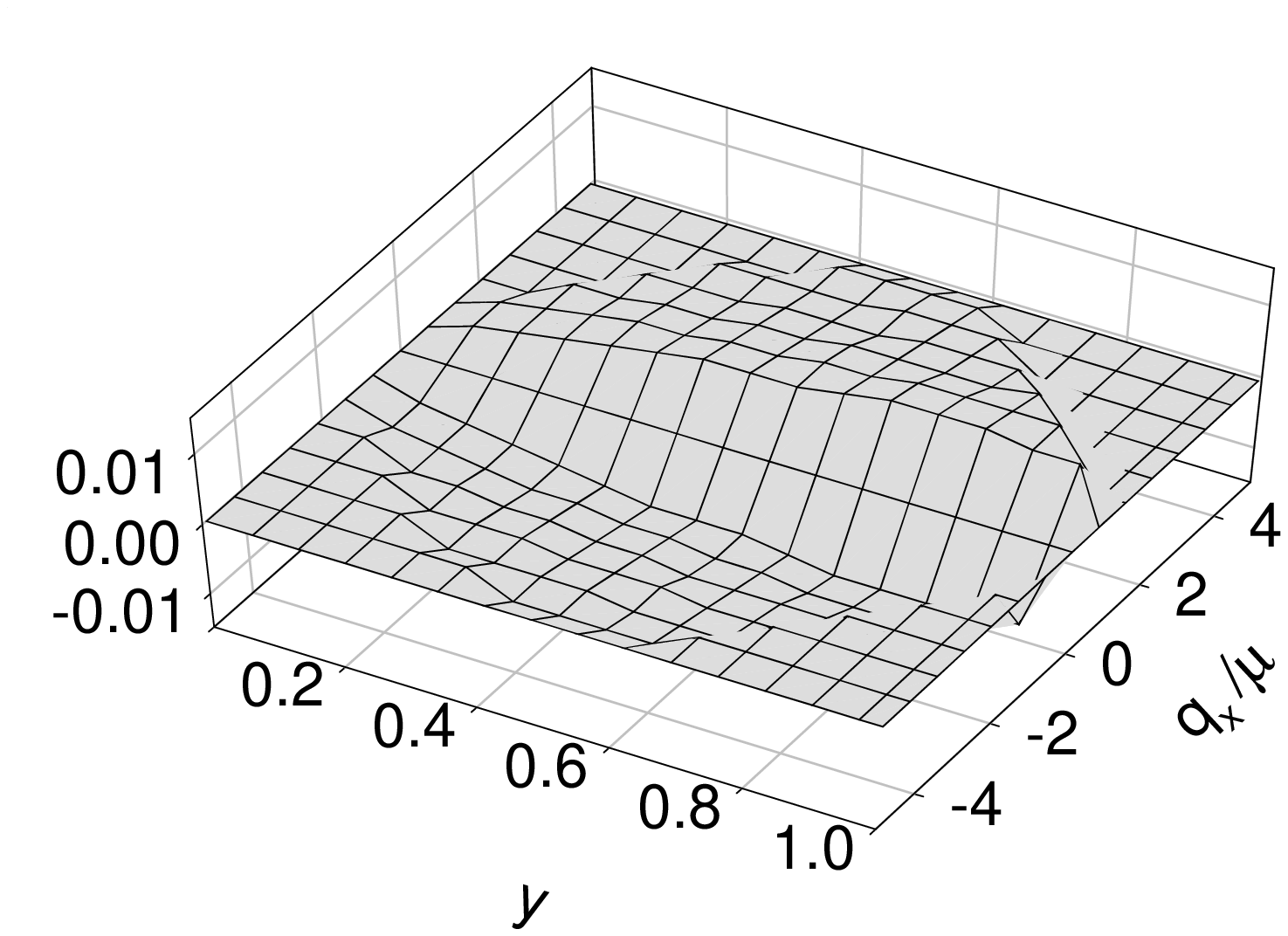,width=3.2in} \\
(a) & (b) \\
\end{tabular}
\caption{\label{fig:phi1000}
The one-boson amplitude $\phi_{\sigma s}^{(1,0,0,0)}$,
with (a) parallel ($s=\sigma$) and 
(b) antiparallel ($s=-\sigma$) bare helicity,
as a function of longitudinal momentum fraction $y$ 
and one transverse momentum component $q_x$ in the
$q_y=0$ plane. The parameter values are 
$K=29$, $N_\perp=7$, $M=\mu$, $\Lambda^2=50\mu^2$,
$\mu_1^2=10\mu^2$, $\mu_2^2=20\mu^2$, $\mu_3^2=30\mu^2$, 
and $\langle:\!\!\phi^2(0)\!\!:\rangle=0.25$. } 
\end{figure}

For higher resolution ($K=21$ and 29 for $\Lambda^2=50\mu^2$ and 
$K=29$ and 39 for $\Lambda^2=100\mu^2$) we plot the structure function
$f_B$ in Figs.~\ref{fig:fBL50} and \ref{fig:fBL100}\@.  Typical
contributions to $f_B$ from one-boson and two-boson states are 
shown in Fig.~\ref{fig:fB100L50onetwo}.
The two-boson contribution to the structure function is further
analyzed in terms of its dependence on both longitudinal momentum
fractions in Fig.~\ref{fig:fBt100L50two}, where
\begin{equation}
\tilde{f}_{B\sigma}(y_1,y_2)\equiv\int d^2q_{\perp 1} d^2 q_{\perp 2} 
    |\phi_{\sigma\sigma}^{(2,0,0,0)}(\underline{q}_j)|^2
\end{equation}
is plotted.  We also show typical two-body wave functions in 
Fig.~\ref{fig:phi1000}; the agreement with 
the necessary $L_z=1$ symmetry in the antiparallel helicity
case is an important check of $J_z$ conservation in the
calculation.

\begin{figure}[htbp]
\centerline{\psfig{file=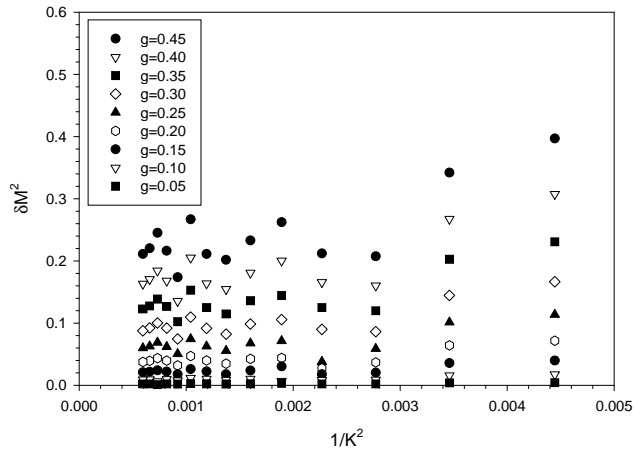,width=3.5in}}
\caption{\label{fig:zeroM2} 
The bare mass shift $\delta M^2$ as a function of 
the longitudinal resolution $K$ for various
bare couplings $g$, with $N_\perp=4$, $M^2=0.001\mu^2$,
$\Lambda^2=50\mu^2$, $\mu_1^2=10\mu^2$, $\mu_2^2=20\mu^2$, and
$\mu_3^2=30\mu^2$. } 
\end{figure}

As a check on the logarithmic PV coupling constraint
in Eq.~(\ref{eq:constraints}), we compute the bare mass
shift $\delta M^2$ when $M^2$ is much less than $\mu^2$.
Figure~\ref{fig:zeroM2} shows its behavior as a function
of the longitudinal resolution for various bare couplings.
If the logarithmic constraint was sufficient for 
nonperturbative calculation, $\delta M^2$ should go
to zero as $M^2$ approaches zero and the resolution
approaches the continuum limit.  This appears to work
well only for weak coupling.

%%%%%%%%%%%%%%%%%%%%%%%%%%%%%%%%%%
\section{Conclusions and Prospects for the Application of DLCQ(3+1) to Gauge Theory}  \label{sec:Conclusions}
%%%%%%%%%%%%%%%%%%%%%%%%%%%%%%%%%%

We have used the discretized light-cone quantization 
method to successfully solve for the mass and light-cone 
wave functions of a dressed fermionic state in a Yukawa 
theory in $3+1$ space-time dimensions. No {\em a priori} 
constraint on the number of bosonic constituents was necessary; 
however, since the fermion constituent was treated as heavy, 
states containing fermion--anti-fermion pairs were truncated.
We have found that the eigensolution at strong coupling displays 
features which  significantly deviated from first-order 
perturbation theory. Numerical resolution in this domain
of strong coupling was not a limiting factor in this analysis.  

The regularization procedure which we have chosen, with three PV
scalars, functioned well.  However, better convergence of the DLCQ 
method at strong coupling could be obtained by constraining the 
PV couplings non-perturbatively, rather than using the one-loop
perturbative constraints in (\ref{eq:constraints}). 

A number of properties of the Yukawa-theory 
eigensolution could be extracted from its light-cone Fock-state  
wave function. This illustrates the power of the DLCQ method
in making the Fock-state wave functions of the eigenstates explicitly
available.  It is also possible to use the derived wave functions
to compute the Pauli and Dirac form factors of the dressed fermion state 
at general momentum transfers~\cite{FormFactors}.

There are additional calculations which  might be done 
within the context of the zero fermion pair approximation.
We could consider two-fermion states and study true bound 
states and scattering solutions.  We could also consider dressed
spin-3/2 states and the analog of N$\pi\leftrightarrow\Delta$
transitions.  Extension of these methods to pseudo-scalar
Yukawa theory would make the N$\pi\leftrightarrow\Delta$
connection stronger.

Although the approach used in this paper to solve (3+1)-dimensional 
Yukawa theory at strong coupling has been successful, future progress 
for solving other quantum field theories will require more 
efficient analytic methods and numerical algorithms.
An alternative ultraviolet regularization procedure, using 
only one PV scalar and one PV fermion~\cite{PastonPrivComm,PV4} 
may potentially provide a more efficient approach for solving 
Yukawa theories.  Not only is the constraint on
the couplings trivial, but  the light-cone Hamiltonian is also
much simpler.  The simplifications occur because the
instantaneous fermion interactions (the terms in (\ref{eq:HLC}) of
order $g^2$) cancel.  Moreover, the DLCQ matrix for the
remaining three-point interactions is much more sparse,
allowing calculations at higher resolutions.
Our next step will be to test this alternative
regularization.

One can consider two other possibilities for PV 
regularization~\cite{PastonPrivComm} of
full Yukawa theory. One is to use two heavy fermions 
and one heavy scalar~\cite{Paston:1999er};
the other is to use one less heavy fermion but make
the transverse momentum cutoff part of the regularization
rather than just a numerical procedure.  In each case
a $\phi^4$ term must be added.  We plan to explore both of these
methods.

Quantum electrodynamics and quantum chromodynamics 
in physical space-time, including the phenomenon of 
chiral symmetry breaking remain  the central challenge 
to DLCQ methods. One attractive possible approach is 
to use broken supersymmetry as an effective ultraviolet 
regulator of the light-cone Hamiltonians of gauge theories. 

The PV method also has applicability to the
renormalization of non-Abelian
Hamiltonian gauge theories on the light-cone.
Paston, Franke and Prokhvatilov~\cite{Paston:1999fq,Paston:2000te} 
have recently extended their analysis to the nonperturbative 
regulation of light-cone QCD, including the regularization 
of the infrared singularities introduced by using light-cone gauge.
They find that a combination of light-cone gauge, PV fields, 
higher derivative regulation, and carefully chosen momentum 
cutoffs can regulate the theory in such a way as
to provide agreement with Feynman calculations using the
Mandelstam--Leibbrandt prescription~\cite{MandelstamLeibbrandt}
for the spurious singularity in the
gauge propagator.  The resulting dynamical operator is 
rather complex and several regulating fields are needed, 
but the regulation procedures appear
suitable for numerical calculations. We plan to 
test these methods in Abelian theory, 
including the calculation of positronium bound 
states~\cite{Krautgartner:1992xz,Kaluza:1992kx}
and  the  non-perturbative calculation of the anomalous 
magnetic moments of leptons at strong coupling~\cite{ae}.

%%%%%%%%%%%%%%%%%%%%%%%%%%%%%%%%%%
\acknowledgments
%%%%%%%%%%%%%%%%%%%%%%%%%%%%%%%%%%
This work was supported in part by the Minnesota Supercomputing Institute
through grants of computing time and by the U.S. Department of Energy,
contracts DE-AC03-76SF00515 (S.J.B.),  DE-FG02-98ER41087 (J.R.H.), and
DE-FG03-95ER40908 (G.M.)\@.

%%%%%%%%%%%%%%%%%%%%%%%%%%%%%%%%%%
\appendix     %start of appendices
%%%%%%%%%%%%%%%%%%%%%%%%%%%%%%%%%%

%%%%%%%%%%%%%%%%%%%%%%%%%%%%%%%%
\section*{Lanczos algorithm for indefinite metric}
\label{sec:Lanczos}
%%%%%%%%%%%%%%%%%%%%%%%%%%%%%%%%

The ordinary Lanczos algorithm~\cite{Lanczos} was designed
for diagonalization of real symmetric or Hermitian matrices.  
A more general form, 
the biorthogonal Lanczos algorithm~\cite{Biorthogonal}, can
be applied to non-symmetric cases but is quite
cumbersome.  In the case of a complex symmetric matrix,
the biorthogonal algorithm can be reduced to a form
nearly as simple as the real symmetric case~\cite{Cullum};
this approach was used in previous work~\cite{PV1,PV2}
where imaginary couplings made negative
norms unnecessary.  The complex-symmetric approach 
is not easy to implement for Yukawa
theory because the Hamiltonian is fully Hermitian.  Instead,
negative norms are assigned, and the eigenvalue problem 
becomes one with indefinite metric.

For this case the biorthogonal algorithm can still be
reduced to a simpler form.  Let $\eta$ represent
the metric signature, so that numerical dot products are written
$\langle\phi'|\phi\rangle={\bbox \phi}^{\,\prime *}\cdot\eta{\bbox \phi}$.
The Hamiltonian matrix $H$ is constructed to be self-adjoint with 
respect to this metric, which means that~\cite{Pauli}
$\bar{H}\equiv\eta^{-1}H^\dagger\eta=H$.  The Lanczos algorithm for
the diagonalization of $H$ then takes the form
\begin{eqnarray} 
\alpha_j&=&\nu_j{\bbox q}_j^{\,*}\cdot\eta H{\bbox q}_j \,, \;\;
   {\bbox r}_j=H{\bbox q}_j-\gamma_{j-1}{\bbox q}_{j-1}
                                          -\alpha_j{\bbox q}_j\,,\;\; 
   \beta_j=+\sqrt{|{\bbox r}_j^{\,*}\cdot\eta{\bbox r}_j|}\,, \;\;
 \\
{\bbox q}_{j+1}&=&{\bbox r}_j/\beta_j\,, \;\;
\nu_{j+1}=\mbox{sign}({\bbox r}_j^{\,*}\cdot\eta{\bbox r}_j)\,,\;\;
   \nu_1=\mbox{sign}({\bbox q}_1^{\,*}\cdot\eta{\bbox q}_1)\,, \;\;
   \gamma_j=\nu_{j+1}\nu_j\beta_j\,,
\nonumber
\end{eqnarray}
where ${\bbox q}_1$ is taken as a normalized initial guess and $\gamma_0=0$.
This initial guess is generated with use of high-order Brillouin--Wigner
perturbation theory.  To determine when to stop the Lanczos iterations, 
the convergence of the eigenvalue and parts of the boson-fermion 
wave function are monitored.

Just as for the ordinary Lanczos algorithm, 
the original matrix $H$ acquires the following tridiagonal matrix 
representation $T$ with respect to the basis formed by the 
vectors ${\bbox q}_j$:
\begin{equation}
H\rightarrow T\equiv\left(\begin{array}{llllll}
              \alpha_1 & \beta_1 & 0 & 0 & 0 & \ldots \\
              \gamma_1 & \alpha_2 & \beta_2 & 0 & 0 & \ldots \\
                0 & \gamma_2 & \alpha_3 & \beta_3 & 0 & \ldots \\
                0 & 0 & \gamma_3 & .  & . & \ldots \\
                0 & 0 & 0 & . & . & \ldots \\
                . & . & . & . & . & \ldots \end{array} \right)\,. 
\end{equation}
By construction, the elements of $T$ are real.  The new matrix is also 
self-adjoint, but with respect to an induced metric $\nu=\{\nu_1,\nu_2,\ldots\}$.
The eigenvalues of $T$ approximate some of the eigenvalues of $H$, even after
only a few iterations.  Approximate eigenvectors of $H$ are constructed
from the right eigenvectors ${\bbox c}_i$ of $T$
as ${\bbox \phi}_i=\sum_k(c_i)_k{\bbox q}_k$.

%%%%%%%%%%%%%%%%%%%%%%%%%%%%%%%%

\end{document}